\documentclass[twocolumn,pra,amsmath,amssymb,superscriptaddress]{revtex4-1}
\usepackage{graphicx,bm,dsfont,amsmath,amssymb}
\usepackage{soul}
\usepackage{color}
\usepackage{amsthm}
\usepackage{mathrsfs}
\usepackage{subfigure}
\usepackage[colorlinks,linkcolor=blue,citecolor=blue,urlcolor=blue]{hyperref}
\begin{document}
	
\title{Synchronization in microwave optomechanical circuits via coupling engineering to a common environment}
\author{Yun-Qiu Ge}
\affiliation{School of Integrated Circuits, Tsinghua University, Beijing 100084, China}
\affiliation{Frontier Science Center for Quantum Information, Beijing, China}
\author{Min-Chen Qiao}
\affiliation{School of Integrated Circuits, Tsinghua University, Beijing 100084, China}
\affiliation{Frontier Science Center for Quantum Information, Beijing, China}
\author{Yu-xi Liu}\email{yuxiliu@mail.tsinghua.edu.cn}
\affiliation{School of Integrated Circuits, Tsinghua University, Beijing 100084, China}
\affiliation{Frontier Science Center for Quantum Information, Beijing, China}

\begin{abstract}
	
Synchronization is one of the essential collective behaviors and has extensive applications. Exploiting a common environment, we establish synchronization in microwave optomechanical circuits. Through analysis and numerical calculations, we study the synchronization dynamics of three nonidentical and mechanically isolated optomechanical resonators. Each resonator supports a microwave mode and a mechanical mode, which are coupled via radiation-pressure-type optomechanical interaction. The common environment induces indirect coupling between any two resonators, which can be described by an effective non-Hermitian interaction Hamiltonian. Combined with the Hermitian interaction regulated by the tunable coupler, we demonstrate that the common environment breaks the reciprocity of the interaction. We propose several special microwave optomechanical circuits with nonreciprocal or even unidirectional interactions, and study the regulation of synchronization dynamics by the common environment. By utilizing the excellent tunability of superconducting circuits, we show that different synchronization states can be switched in a controllable way. This work may open up a new way for synchronization research and have potential applications in synchronization networks.

\end{abstract}

\maketitle

\section{Introduction}

Synchronization phenomena are ubiquitous in coupled-oscillator systems~\cite{SynchronizationBook}. Their essence lies in the adjustment of the rhythms in oscillators due to weak interactions~\cite{chaos2000,nature2001,PhysRevApplied.12.054039}. Since Huygens observed the synchronized oscillations of pendulum clocks~\cite{PhysRep.517.1,scirep.2016}, the research on synchronization has attracted widespread attention, e.g., the synchronization of neurons~\cite{science3162007,nature2016}, digital telecommunications networks~\cite{SynchronizationBook2}, lasers~\cite{nature2023,PRL722009}, and chemical oscillators~\cite{science3232009,RevModPhys.77.2005}. Over the past few decades, numerous types of synchronization behaviors have been recognized, e.g., lag synchronization~\cite{PRL784193} and chaotic synchronization~\cite{PhysRep.366.1}. The corresponding methods for achieving synchronization have also been greatly enriched. Representative examples include unidirectional or bidirectional coupling~\cite{PRL108214101,Photonic92021,OptExpress.24.12336}, correlated noise~\cite{PRL88230602,PRL128098301}, and external drive~\cite{pra952017,PRL112094102}. Benefiting from the maturity of micro- and nano-fabrication techniques~\cite{fabricationbook}, the size of various optical, electrical, and mechanical oscillators, as well as their hybrid devices, has reached the micron level or even smaller~\cite{PhysRep.718.1}. This makes it easy to realize on-chip synchronization for real-time communications, accurate timekeeping, and advanced navigation~\cite{PhysRep.469.93,RMP85623}.

Microwave optomechanical resonators are miniaturized devices at the micron level with good scalability and high controllability~\cite{RMP861391,naturephysics2022}. Thus, it is of great interest to achieve on-chip synchronization by utilizing microwave optomechanical resonators. By coupling two or more resonators, one can construct microwave optomechanical circuits for the observation, control, and application of synchronization~\cite{PRL111073603,PRL107043603,PRL111084101,PRA2018023841}. In experiments, the interaction between microwave optomechanical resonators can be achieved via a variety of methods, e.g., coplanar waveguide or transmission line~\cite{PRL125023603}, mutual inductance~\cite{nature6662022}, capacitor~\cite{PRX2017031001}, or mechanical link~\cite{PRL123017402}. It should be noted that any system inevitably interacts with its surrounding environment. Although the environmental effects in microwave optomechanical circuits have been studied from many perspectives~\cite{naturephysics2022,PhysRep.718.1}, the synchronization induced by the environment remains less explored.

In this paper, we will study the environment-induced synchronization in microwave optomechanical circuits composed of multiple optomechanical resonators. Based on the environmental effects on single and multiple subsystems of a composite system, the environments can be classified into two types: one is independent or uncorrelated environment~\cite{En12019,naturephysics172021}, and the other is common or correlated environment~\cite{En2,PhysRevX.5.021025}. Thus, in microwave circuits constructed by multiple optomechanical resonators, the environment may take the following three forms to affect each subsystem (i.e., an optomechanical resonator). (i) Each subsystem interacts only with its own independent environment, which has no effect on the other subsystems. (ii) Multiple or all subsystems share one or more common environments. (iii) Each subsystem not only interacts with its independent environment, but also shares common environments with other subsystems. For the entire circuit, it has been shown that the independent environment for each subsystem cannot result in the coupling between different subsystems~\cite{RMP882016021002,RMP892017015001}, while the common environment, e.g., substrate materials, microwave waveguides, or transmission lines, shared by multiple subsystems can induce indirect couplings between subsystems~\cite{PRL121Harder,PhysRevX.5.021025}. Therefore, the common environment may result in synchronized oscillations of optomechanical resonators. We emphasize that the indirect couplings induced by the common environment are non-Hermitian~\cite{PhysRevLett.123.127202,PhysRevLett.127.250402,PhysRevX.5.021025}. This feature may bring intriguing implications for synchronization phenomena, especially when there are Hermitian couplings between resonators. We will also study the relation between non-Hermitian couplings and synchronization.

This paper is organized as follows. In Sec.~\ref{sec2}, a theoretical Hamiltonian is proposed to describe the model of microwave optomechanical circuits with a common environment. The effective non-Hermitian interaction Hamiltonian between optomechanical resonators mediated by the common environment is presented. In Sec.~\ref{sec3}, we study the synchronization dynamics in the circuit induced by the common environment utilizing the Heisenberg equations of motion. In Sec.~\ref{sec4}, we construct several special microwave optomechanical circuits by combining the non-Hermitian couplings with the Hermitian couplings between resonators. The synchronization dynamics exhibited in these circuits are further studied. Finally, in Sec.~\ref{sec5}, we summarize the main results and discuss the experimental feasibility of our proposal.

\section{Theoretical model based on microwave optomechanical circuits \label{sec2}}

\begin{figure}
	\centering
	\includegraphics[width=8.5cm]{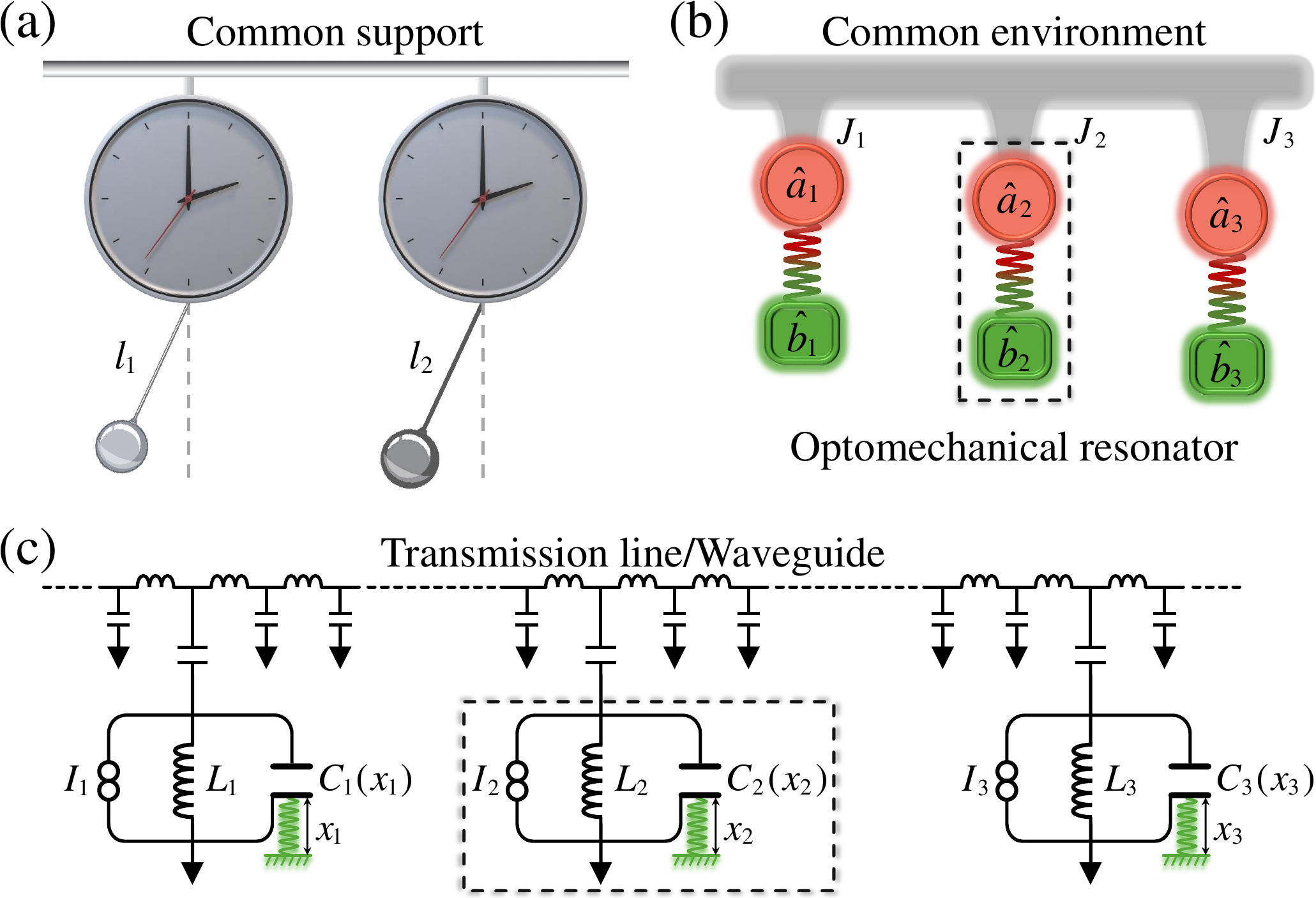}
	\caption{(a) Schematic diagram of the experiment designed by Huygens to synchronize nonidentical pendulum clocks, where the two pendulum clocks with pendulum lengths $l_{1}$ and $l_{2}$ are placed on a common support. (b) Schematic diagram of three nonidentical microwave optomechanical resonators coupled to a common environment, where the $j$-th resonator supports a microwave mode $\hat{a}_{j}$ and a mechanical mode $\hat{b}_{j}$, $j=1,2,3$. (c) Implementation of the equivalent circuit, where the common environment is simulated by a transmission line or waveguide. The black dashed box represents the equivalent circuit of a microwave optomechanical resonator, which includes a linear inductor $L_{2}$ and a vibrating capacitor $C_{2}(x_{2})$. Here, $x_{2}$ represents the distance between the upper and lower plates, and varies with the vibrations of one of the plates. The resonator can be pumped via an external current $I_{2}$.}
	\label{Fig1}
	\vspace{-0.3cm}
\end{figure}

One of the famous experiments to observe and explore synchronization phenomena is Huygens' pendulum clock experiment, which is presented in Fig.~\ref{Fig1}(a). Through the indirect interaction mediated by a common support, two nonidentical pendulum clocks ($l_{1}\neq\l_{2}$) can oscillate with an identical frequency~\cite{SynchronizationBook,PhysRep.517.1}. Inspired by Huygens' experiment, we study the synchronization dynamics of three nonidentical microwave optomechanical resonators, which are coupled to a common environment, as schematically shown in Figs.~\ref{Fig1}(b) and \ref{Fig1}(c). The $j$-th microwave optomechanical resonator in the circuit (the black dashed box in Fig.~\ref{Fig1}) is formed by a linear inductor $L_{j}$ and a vibrating capacitor $C_{j}(x_{j})$, in which $x_{j}=x_{j,0}+\delta x_{j}$ ($\delta x_{j}\ll x_{j,0}$) represents the distance between the upper and lower plates of the capacitor~\cite{Nature2011204471}. Here, $x_{j,0}$ denotes the distance between the two plates when they are in equilibrium positions, and $\delta x_{j}$ denotes the small change of the distance due to the vibration of one of the plates. Such an optomechanical resonator supports a microwave mode $\hat{a}_{j}$ with a resonance frequency $\omega_{j}=1/\sqrt{L_{j}C_{j}(x_{j,0})}$ and a mechanical mode $\hat{b}_{j}$ with a resonance frequency $\Omega_{j}$, $j=1,2,3$. To better demonstrate the effects of the common environment on the synchronization behaviors of these resonators, we first consider the case where there is no direct interaction between resonators. More complex cases will be discussed in Sec.~\ref{sec4}. The Hamiltonian of the microwave optomechanical circuit without the common environment is given by (henceforth, we set $\hbar=1$)
\begin{equation}
	\hat{H}_{\rm OM}=\sum_{j=1}^{3}\omega_{j}\hat{a}_{j}^{\dag}\hat{a}_{j}+\Omega_{j}\hat{b}_{j}^{\dag}\hat{b}_{j}-G_{j}\hat{a}_{j}^{\dag}\hat{a}_{j}(\hat{b}_{j}^{\dag}+\hat{b}_{j}),
	\label{Eq1}
\end{equation}
in which $G_{j}$ represents the single-photon optomechanical coupling strength of the $j$-th resonator.

We assume that the three optomechanical resonators not only have their own independent intrinsic environments but also share a common environment. In actual circuits, the common environment may consist of substrate materials, microwave waveguides, or transmission lines~\cite{PhysRep.718.1,PhysRevX.5.021025}. We here assume that the common environment is a transmission line, as schematically shown in Fig.~\ref{Fig1}(c). Then the Hamiltonian of the common environment is
\begin{equation}
	\hat{H}_{\rm E}=\int_{0}^{\infty}\!d\omega\;\;\!\!\!\omega\!\left[\hat{c}_{\rm L}^{\dag}(\omega)\hat{c}_{\rm L}(\omega)+\hat{c}_{\rm R}^{\dag}(\omega)\hat{c}_{\rm R}(\omega)\right]\!\;\!\!,
	\label{Eq2}
\end{equation}
where $\hat{c}_{\rm L(R)}(\omega)$ and $\hat{c}_{\rm L(R)}^{\dag}(\omega)$ represent the annihilation and creation operators of the left (right) moving environmental mode at frequency $\omega$. The optomechanical resonators are capacitively coupled to the transmission line~\cite{PhysRep.718.1,RMP85623}, thus, the interaction Hamiltonian between the common environment and the optomechanical resonators can be described as~\cite{PhysRep.718.1,PRA88043806}
\begin{equation}
	\begin{split}
		\hat{H}_{\rm I}=&\;\;\!\!\!\sum_{j=1}^{3}\int_{0}^{\infty}\!d\omega J_{j}
		\sqrt{\omega}\;\;\!\!\!\hat{a}_{j}^{\dag}\!\left[\hat{c}_{\rm L}(\omega)e^{-i\omega x_{j}/v}+\hat{c}_{\rm R}(\omega)e^{i\omega x_{j}/v}\right] \\
		+&\;\;\!\!\!\sum_{j=1}^{3}\int_{0}^{\infty}\!d\omega J_{j}
		\sqrt{\omega}\;\;\!\!\!\hat{a}_{j}\!\left[\hat{c}_{\rm L}^{\dag}(\omega)e^{i\omega x_{j}/v}+\hat{c}_{\rm R}^{\dag}(\omega)e^{-i\omega x_{j}/v}\right]\!\;\!\!,
		\label{Eq3}
	\end{split}
\end{equation}
where $x_{j}$ is the position of the $j$-th resonator, $J_{j}$ depends on the position of the $j$-th optomechanical resonator and is proportional to the coupling strength between the $j$-th resonator and the common environment. The parameter $v$ represents the group velocity (linear dispersion relation \cite{PRA042116}) corresponding to the environmental modes.

To facilitate discussion, we assume that the frequencies of the three microwave optomechanical resonators satisfy $\omega_{1}=\omega_{0}-\delta$, $\omega_{2}=\omega_{0}$, $\omega_{3}=\omega_{0}+\delta$, $\Omega_{1}=\Omega_{0}-\Delta$, $\Omega_{2}=\Omega_{0}$, $\Omega_{0}=2\pi\times10\;{\rm MHz}$, and $\Omega_{3}=\Omega_{0}+\Delta$. Therefore, as shown in Appendix~\ref{Appendix1}, by eliminating the degrees of freedom of the common environment, we can obtain the effective non-Hermitian Hamiltonian between resonators from Eqs.~(\ref{Eq2}) and (\ref{Eq3}) as follows
\begin{align}
	\hat{H}_{\rm N}=&-i2\pi\;\;\!\!\!\omega_{0}\sum_{j=1}^{3}J_{j}^{2}\hat{a}^{\dagger}_{j}\hat{a}_{j}\nonumber\\
	&-i2\pi J_{1}J_{2}\;\;\!\!\!\omega_{0}(\hat{a}_{2}^{\dagger}\hat{a}_{1}e^{i\theta}+\hat{a}_{1}^{\dagger}\hat{a}_{2}e^{i\theta})\nonumber\\
	&-i2\pi J_{2}J_{3}\;\;\!\!\!\omega_{0}(\hat{a}_{3}^{\dagger}\hat{a}_{2}e^{i\varphi}+\hat{a}_{2}^{\dagger}\hat{a}_{3}e^{i\varphi})\nonumber\\
	&-i2\pi J_{1}J_{3}\;\;\!\!\!\omega_{0}\;\!\!\Big[\hat{a}_{3}^{\dagger}\hat{a}_{1}e^{i(\theta+\varphi)}+\hat{a}_{1}^{\dagger}\hat{a}_{3}e^{i(\theta+\varphi)}\Big]\;\!\!,
	\label{Eq4}
\end{align}
in which the phase parameters $\theta$ and $\varphi$ are defined as
\begin{align}
    \theta=\omega_{0}|x_{2}-x_{1}|/v,\quad \varphi=\omega_{0}|x_{3}-x_{2}|/v.
	\label{Eq5}
\end{align}
Equation~(\ref{Eq4}) shows that the common environment not only results in the on-site dissipation of each resonator, but also induces dissipative couplings between different resonators. The on-site dissipations are characterized by the terms with $J_{j}^2$ for $j=1,2,3$. Other terms represent dissipative couplings, which include both Hermitian and non-Hermitian parts. The Hermitian coupling strengths between $a_{1}$ and $a_{2}$, $a_{2}$ and $a_{3}$, and $a_{3}$ and $a_{1}$ are $2\pi J_{1}J_{2}\;\;\!\!\!\omega_{0}\;\;\!\!\!{\rm sin}\;\;\!\!\!\theta$, $2\pi J_{2}J_{3}\;\;\!\!\!\omega_{0}\;\;\!\!\!{\rm sin}\;\;\!\!\!\varphi$, and $2\pi J_{1}J_{3}\;\;\!\!\!\omega_{0}\;\;\!\!\!{\rm sin}(\theta+\varphi)$, respectively. The corresponding non-Hermitian coupling strengths between $a_{1}$ and $a_{2}$, $a_{2}$ and $a_{3}$, and $a_{3}$ and $a_{1}$ are $2\pi J_{1}J_{2}\;\;\!\!\!\omega_{0}\;\;\!\!\!{\rm cos}\;\;\!\!\!\theta$, $2\pi J_{2}J_{3}\;\;\!\!\!\omega_{0}\;\;\!\!\!{\rm cos}\;\;\!\!\!\varphi$, and $2\pi J_{1}J_{3}\;\;\!\!\!\omega_{0}\;\;\!\!\!{\rm cos}\;\;\!\!\!(\theta+\varphi)$, respectively.

Based on the effective Hamiltonian of the entire circuit, i.e., $\hat{H}_{\rm C}^{\rm eff}=\hat{H}_{\rm OM}+\hat{H}_{\rm N}$, as shown in Eqs.~(\ref{Eq1}) and (\ref{Eq4}), and considering the independent environments for all microwave and mechanical modes, then the Heisenberg-Langevin equations of motion for the microwave and mechanical modes can be expressed as
\begin{align}
	\dfrac{da_{1}}{dt}=&\left[i\delta-\gamma_{1}-2\pi J_{1}^2\omega_{0}+iG(b_{1}+b_{1}^*)\right]\!a_{1}+\varepsilon \nonumber\\
	-&\:2\pi J_{1}J_{2}\;\;\!\!\!\omega_{0}e^{i\theta}a_{2}-2\pi J_{1}J_{3}\;\;\!\!\!\omega_{0}e^{i(\theta+\varphi)}a_{3},\label{Eq6}\\
	\dfrac{da_{2}}{dt}=&\left[-\:\gamma_{2}-2\pi J_{2}^2\omega_{0}+iG(b_{2}+b_{2}^*)\right]\!a_{2}+\varepsilon \nonumber\\
	-&\:2\pi J_{1}J_{2}\;\;\!\!\!\omega_{0}e^{i\theta}a_{1}-2\pi J_{2}J_{3}\;\;\!\!\!\omega_{0}e^{i\varphi}a_{3},\label{Eq7}\\
	\dfrac{da_{3}}{dt}=&\left[-\:i\delta-\gamma_{3}-2\pi J_{3}^2\omega_{0}+iG(b_{3}+b_{3}^*)\right]\!a_{3}+\varepsilon \nonumber\\-&\:2\pi J_{1}J_{3}\;\;\!\!\!\omega_{0}e^{i(\theta+\varphi)}a_{1}-2\pi J_{2}J_{3}\;\;\!\!\!\omega_{0}e^{i\varphi}a_{2},\label{Eq8}
\end{align}	
\begin{align}	
	\dfrac{db_{\rm 1}}{dt}=&\;\!\;\!\!\left[-\:i(\Omega_{0}-\Delta)-\Gamma_{1}
	\right]\:\!\!b_{\rm 1}+iG|a_{1}|^2,\label{Eq9}\\[2pt]	
	\dfrac{db_{\rm 2}}{dt}=&\;\;\!\!\!\left(-\:i\Omega_{0}-\Gamma_{2}\right)\:\!\!b_{\rm 2}+iG|a_{2}|^2,\label{Eq10}\\[2pt]
	\dfrac{db_{\rm 3}}{dt}=&\;\;\!\!\!\left[-\:i(\Omega_{0}+\Delta)-\Gamma_{3}\right]\:\!\!b_{\rm 3}+iG|a_{3}|^2.
	\label{Eq11}
\end{align}
Here, the damping rates $\gamma_{j}$ and $\Gamma_{j}$ of the modes $\hat{a}_{j}$ and $\hat{b}_{j}$ originate from their own independent environments, respectively. However, the damping rate $2\pi J_{j}^2\omega_{0}$ for the mode $\hat{a}_{j}$ with $j=1,2,3$ is induced by the common environment. We note that the following assumptions and approximations are made in deriving Eqs.~(\ref{Eq6})--(\ref{Eq11}). (i) We assume that an ac current with frequency $\omega_{0}$ and amplitude $\varepsilon$ is applied to each microwave mode~\cite{PhysRevLett.131.067001,PhysRevLett.115.233601,PRL103213603}. (ii) We work in a rotating reference frame with the frequency $\omega_{0}$ of the driving field. (iii) We apply the mean field approximation and neglect the second- and higher-order correlations of the microwave-microwave modes and the microwave-mechanical modes~\cite{PRL111073603,PRL103213603}. That is, the operators have been replaced with their expectation values, specifically, $a_{j}e^{-i\omega_{0}t}=\langle\hat{a}_{j}\rangle$ and $b_{j}=\langle\hat{b}_{j}\rangle$. Here, $a_{j}$ and $b_{j}$ denote the amplitudes of the microwave and mechanical modes in the $j$-th optomechanical resonator, respectively. Moreover, we emphasize that Eqs.~(\ref{Eq6})--(\ref{Eq11}) can also be derived directly from the Hamiltonian $\hat{H}_{\rm C}=\hat{H}_{\rm  OM}+\hat{H}_{\rm E}+\hat{H}_{\rm I}$ through the Heisenberg-Langevin equations of motion (for more details, see Appendix~\ref{Appendix2}). To facilitate the subsequent discussions, we assume that $2\pi J_{1}J_{2}\;\;\!\!\!\omega_{0}=2\pi J_{2}J_{3}\;\;\!\!\!\omega_{0}=2\pi J_{1}J_{3}\;\;\!\!\!\omega_{0}=J$, $\gamma_{1}+2\pi J_{1}^2\omega_{0}=\gamma_{2}+2\pi J_{2}^2\omega_{0}=\gamma_{3}+2\pi J_{3}^2\omega_{0}=\gamma$, and $\Gamma_{1}=\Gamma_{2}=\Gamma_{3}=\Gamma$. 

\section{Synchronization via common environment \label{sec3}}

\begin{figure*}
	\centering
	\includegraphics[width=16.5cm]{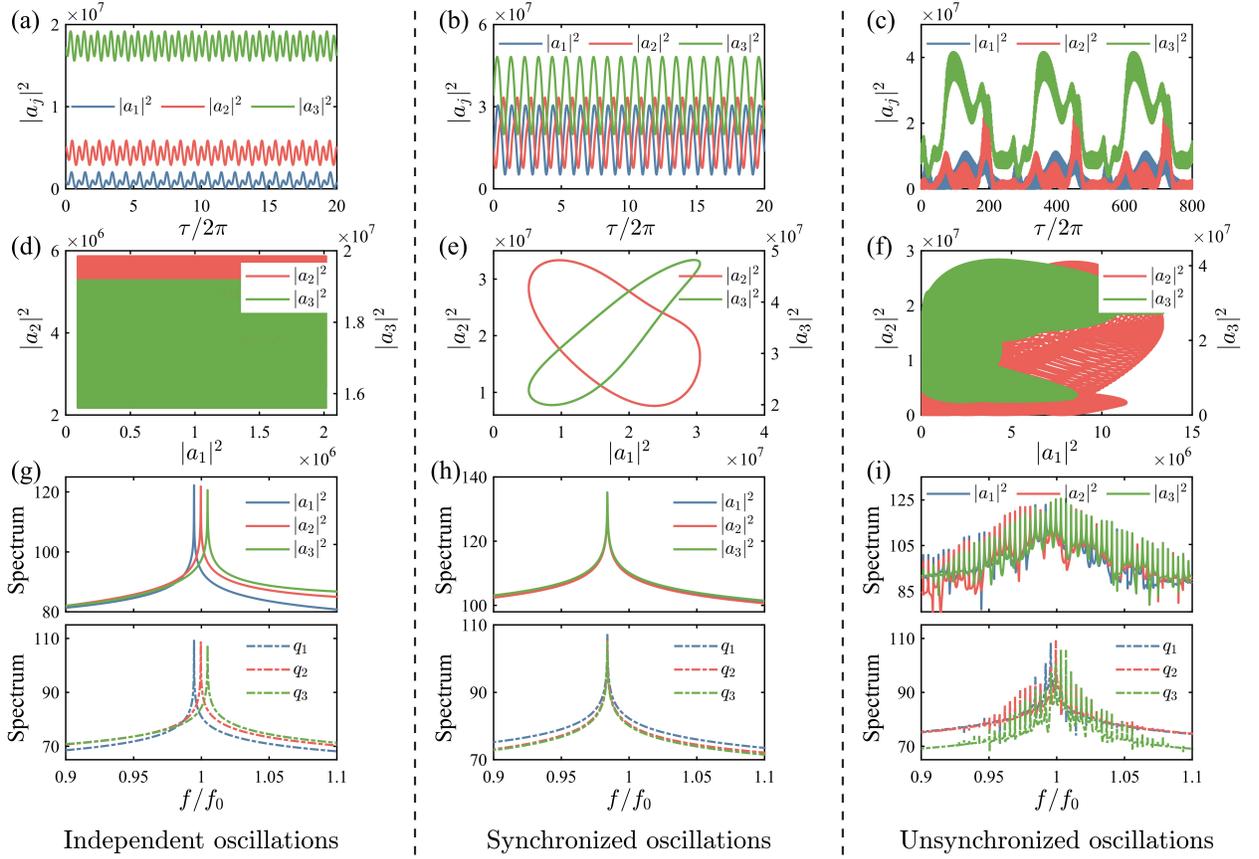}
	\caption{(a)--(c) Time-domain evolution of the intensity (i.e., $|a_{j}|^2$) of microwave modes in the circuit, under different parameter conditions. (d)--(f) The corresponding trajectories of $|a_{1}|^2,\,|a_{2}|^2$ (red curves) and $|a_{1}|^2,\,|a_{3}|^2$ (green curves). (g)--(i) Frequency spectra of microwave modes $|a_{j}|^2$ (upper panels) and mechanical modes $q_{j}$ (lower panels) for different oscillation states. In this figure, blue, red, and green solid curves are associated with $|a_{1}|^2$, $|a_{2}|^2$, and $|a_{3}|^2$, respectively; blue, red, and green dash-dotted curves are associated with $q_{1}$, $q_{2}$, and $q_{3}$, respectively. We define $q_{j}={\rm Re}(b_{j}\;\!\!)$ as the displacement of the $j$-th mechanical mode. Parameters: $\tau=\Omega_{0}t$, $\Omega_{0}=2\pi\;\!\! f_{0}$, $\delta/\Omega_{0}=0.05$, $\Delta/\Omega_{0}=5\times10^{-3}$, $G/\Omega_{0}=4\times10^{-5}$, $\varepsilon/\Omega_{0}=800$, $\gamma/\Omega_{0}=0.1$, $\Gamma/\Omega_{0}=8\times10^{-5}$, in (a),(d),(g), $J=0$, in (b),(e),(h), $J/\Omega_{0}=0.11$, $\theta=\varphi=0.5\pi$, in (c),(f),(i), $J/\Omega_{0}=0.11$, $\theta=0.6\pi$, $\varphi=0.8\pi$~\cite{RMP861391,naturephysics2022}.}
	\label{Fig2}
	\vspace{-0.3cm}
\end{figure*}

We now numerically solve the equations of motion in Eqs.~(\ref{Eq6})--(\ref{Eq11}) to observe the synchronization behavior of the optomechanical resonators in the circuit. Figure~\ref{Fig2} depicts several representative oscillation states, including independent oscillations, unsynchronized oscillations, and synchronized oscillations. We here emphasize that the initial transient processes have been discarded in all figures. As shown in Figs.~\ref{Fig2}(a) and~\ref{Fig2}(d), if the common environment does not exist, i.e., $J=0$, then the three optomechanical resonators oscillate independently, and the amplitudes of self-induced oscillations are widely separated. Frequency spectra with three peaks (equally spaced with $\Delta$) in Fig.~\ref{Fig2}(g) also confirm this oscillation state. When the common environment exists, such as $J/\Omega_{0}=0.11$, $\theta=0.5\pi$, and $\varphi=0.5\pi$, the situation is completely different. The time-domain evolution of $|a_{j}|^2$ presented in Fig.~\ref{Fig2}(b) indicates that the rhythms in the resonators are tuned and a collective oscillatory behavior emerges. Moreover, the smooth trajectories in Fig.~\ref{Fig2}(e) confirm the state of synchronized oscillations~\cite{PRL129063605}. The single peak in the spectra of microwave modes $|a_{j}|^2$ (upper panel) and mechanical modes $q_{j}$ (lower panel) in Fig.~\ref{Fig2}(h) demonstrates the frequency synchronization phenomenon induced by the common environment. $q_{j}$ is defined as $q_{j}={\rm Re}(b_{j}\;\!\!)$, which represents the displacement of the mechanical mode $b_{j}$.

It should be noted that the phases $\theta$ and $\varphi$ also play an important role in the synchronization dynamics when $J\neq0$. For example, we depict Figs.~\ref{Fig2}(c) and~\ref{Fig2}(f) with the same coupling strength $J/\Omega_{0}=0.11$ as in Figs.~\ref{Fig2}(b) and~\ref{Fig2}(e), but with different phases $\theta=0.6\pi$ and $\varphi=0.8\pi$. Figures~\ref{Fig2}(c) and~\ref{Fig2}(f) clearly show that the coupling strength $J$ cannot solely determine the synchronization dynamics of the proposed circuit. Although the rhythms in resonators are adjusted, there is no obvious correlation or collectiveness between the oscillatory behaviors for $\theta=0.6\pi$ and $\varphi=0.8\pi$ even with the same coupling strength $J$ as in Figs.~\ref{Fig2}(b) and~\ref{Fig2}(e). Moreover, we can observe a series of sideband peaks from the spectra of $|a_{j}|^2$ and $q_{j}$, as shown in Fig.~\ref{Fig2}(i). All these phenomena indicate that the oscillations are unsynchronized. Thus, we conclude that the effects of parameters $\theta$ and $\varphi$ on the dynamics of optomechanical circuits are significant.

\begin{figure*}[t]
	\centering
	\includegraphics[width=16cm]{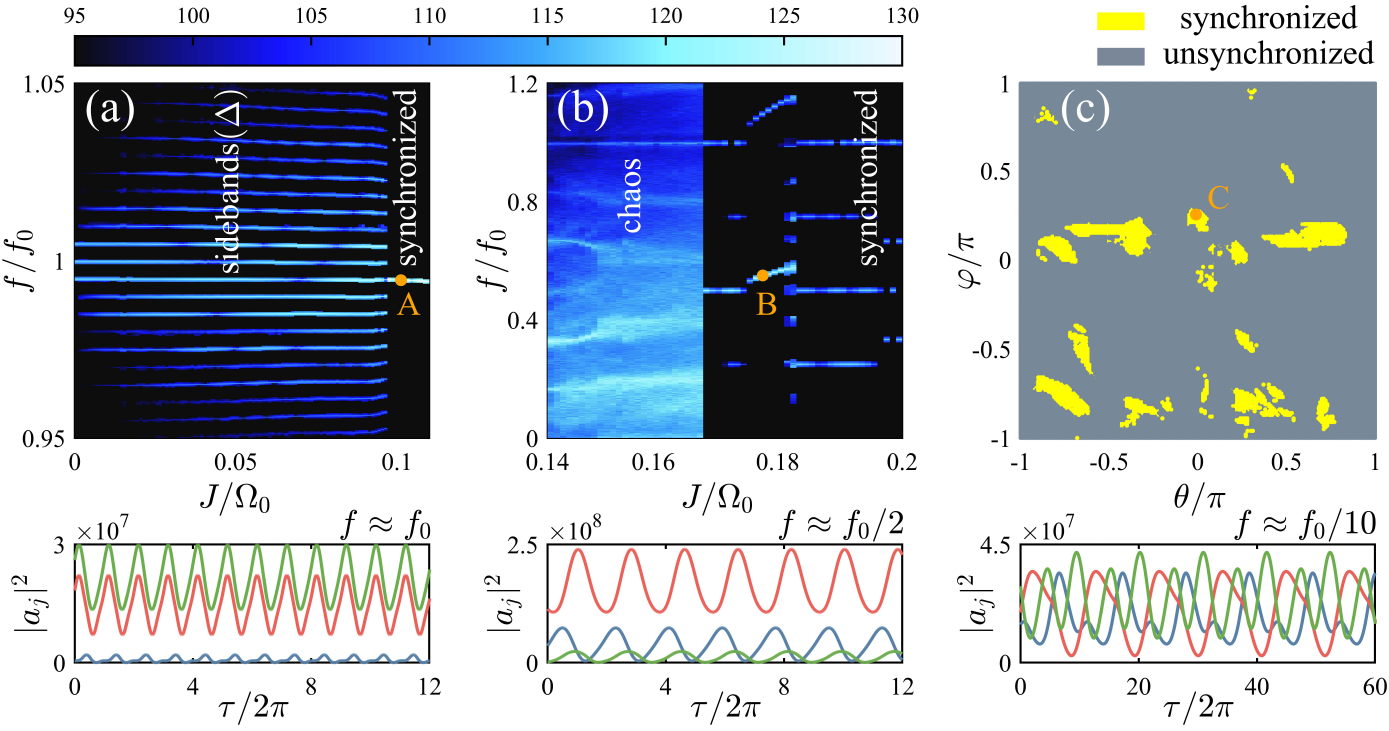}
	\caption{Frequency spectra of the entire circuit versus the coupling strength $J$ in (a) for $J/\Omega_{0}\in[0, 0.11]$ and in (b) for $J/\Omega_{0}\in[0.14, 0.2]$. (c) Phase diagram of the system within the $(\theta,\varphi)$ plane. We here only highlight the frequency synchronization, i.e., the yellow region, and all other phases are classified as unsynchronization, i.e., the grey region. The three panels at the bottom present the time domain evolution of $|a_{j}|^2$ for three representative points A, B, and C, respectively. Blue, red, and green solid curves represent $|a_{1}|^2$, $|a_{2}|^2$, and $|a_{3}|^2$, respectively. Parameters in (a) and (b) are taken as $\theta=0.2\pi$, $\varphi=-\;\!0.8\pi$. The point A in (a) corresponds to $J/\Omega_{0}=0.1$; the point B in (b) corresponds to $J/\Omega_{0}=0.18$. The parameter in (c) is taken as $J/\Omega_{0}=0.11$, and the point C corresponds to $\theta=0$ and $\varphi=0.2\pi$. The selection of other parameters is the same as that in Fig.~\ref{Fig2}.}
	\label{Fig3}
	\vspace{-0.3cm}
\end{figure*}

To illustrate how $J$, $\theta$, and $\varphi$ regulate the synchronization behavior, we present Fig.~\ref{Fig3} with different $J$, $\theta$, and $\varphi$. As shown in the frequency spectra in Fig.~\ref{Fig3}(a), when the coupling strength $J$ is weak, the rhythms (each line in the spectra represents a rhythm) in the system hardly change. As $J$ increases, the original three rhythms begin to mix, generating a series of higher-order sidebands with frequency difference $\Delta$. Only when the coupling strength exceeds a critical threshold, the synchronization effect dominates other nonlinear effects, and the synchronized oscillations with a single peak in the frequency spectrum occur. It should be noted that due to the nonlinearity of optomechanical interactions, the system can exhibit very rich nonlinear dynamics, which compete with each other \cite{Nature202175,NatPhoton103992016,NC1115892020}. Thus, as shown in Fig.~\ref{Fig3}(b), the synchronization region is not continuous, but intertwined with some other regions, e.g., chaotic region. Moreover, the frequencies of synchronized oscillations can vary significantly. From the time-domain evolution of $|a_{j}|^2$ corresponding to points A and B, we find that the synchronization frequency of B ($f\approx f_{0}/2$) is approximately half that of A ($f\approx f_{0}$). This phenomenon indicates that the common environment can induce multiple synchronization dynamics in different frequency regimes. Figure~\ref{Fig3}(c) shows the phase diagram on the $(\theta,\varphi)$ plane for a fixed $J$. By regulating the equivalent distance between microwave optomechanical resonators ($\theta=\omega_{0}|x_{2}-x_{1}|/v$ and $\varphi=\omega_{0}|x_{3}-x_{2}|/v$), the oscillation state of the circuit can be switched between synchronized (the yellow region) and unsynchronized (the grey region). The synchronized frequency $f$ can also be tuned to be much lower than the original frequencies, as shown in the bottom of Fig.~\ref{Fig3}(c), where $f\approx f_{0}/10$ for the representative point C. Thus, the synchronization of optomechanical resonators can be regulated by varying the parameters $J$, $\theta$, and $\varphi$, which characterize the couplings between the optomechanical resonators and the common environment. This can be achieved by adjusting the arrangement of resonators or equivalently tuning the interactions between optomechanical resonators and the transmission line~\cite{PhysRep.718.1}.

Physically, the common environment carries oscillation information from all three nonidentical optomechanical resonators (each one is in self-induced oscillation) and transmits information bidirectionally between all the resonators. Thus, the rhythms of resonators are tuned and synchronization phenomena can occur. However, due to the nonlinearity of optomechanical interactions, after the three resonators are synchronized to a certain frequency $f$, if we further adjust the parameters, period-doubling bifurcation cascades may be induced~\cite{M.C.Gutzwiller,NatPhoton9151,PRL1140136012015,chiralchaos}. This means that the synchronization frequency can be reduced to $f/n$, where $n$ denotes a positive integer. This is the origin of the multiple synchronization frequencies (e.g., $f\approx f_{0}$, $f_{0}/2$, and $f_{0}/10$) established in Fig.~\ref{Fig3}. We focus on frequency synchronization in this paper, while other types of synchronization will be explored in future work.

\section{Synchronization dynamics regulated by common environment \label{sec4}}

We now consider the more general case, where there exists direct tunable coupling in addition to the indirect coupling induced by the common environment between any two microwave optomechanical resonators. Here, the direct tunable coupling can be realized via various couplers, which have been proposed and implemented in superconducting microwave circuits~\cite{NewJPhys.10.115001,PhysRevLett.113.220502,PhysRevLett.112.123601,APR62019,PRA2018023841,PhysRevX.3.021013,RMP93025005,RMP91025005} with large tunable range~\cite{RMP85623} and high control accuracy~\cite{PhysRep.718.1}. The interaction Hamiltonian between resonators through couplers can be written as
\begin{equation}
	\hat{H}_{0}=g_{1}e^{i\phi_{1}}\hat{a}_{1}^{\dag}\hat{a}_{2}+g_{2}e^{i\phi_{2}}\hat{a}_{2}^{\dag}\hat{a}_{3}+g_{3}e^{i\phi_{3}}\hat{a}_{3}^{\dag}\hat{a}_{1}+\rm H.c.,
	\label{Eq12}
\end{equation}
where $g_{j}$ with $j=1,2,3$ represents the coherent coupling strength between resonators. We define $\hat{H}_{0}$ as the coherent interaction Hamiltonian, which makes the system evolve unitarily. The parameters $\phi_{1}$, $\phi_{2}$, and $\phi_{3}$ represent the phases corresponding to the coupling strengths $g_{1}$, $g_{2}$, and $g_{3}$, respectively. Combined with the dissipative couplings in Eq.~(\ref{Eq4}) induced by the common environment, the effective interaction Hamiltonian of the entire system can be written as
\begin{equation}
	\begin{split}	\hat{H}&=\left(g_{1}e^{i\phi_{1}}-iJe^{i\theta}\right)\!\hat{a}_{1}^{\dag}\hat{a}_{2}+\left(g_{1}e^{-i\phi_{1}}-iJe^{i\theta}\right)\!\hat{a}_{2}^{\dag}\hat{a}_{1}\\&+\left(g_{2}e^{i\phi_{2}}-iJe^{i\varphi}\right)\!\hat{a}_{2}^{\dag}\hat{a}_{3}+\left(g_{2}e^{-i\phi_{2}}-iJe^{i\varphi}\right)\!\hat{a}_{3}^{\dag}\hat{a}_{2}\\
	&+\!\;\!\Big[g_{3}e^{i\phi_{3}}-iJe^{i(\theta+\varphi)}\Big]\;\!\!\hat{a}_{3}^{\dag}\hat{a}_{1}+\Big[g_{3}e^{-i\phi_{3}}-iJe^{i(\theta+\varphi)}\Big]\!\;\!\hat{a}_{1}^{\dag}\hat{a}_{3}.
	\label{Eq13}
	\end{split}
\end{equation}
Hereafter, the coupling with parameter $J$ induced by the common environment is defined as dissipative coupling. Equation~(\ref{Eq13}) shows that the common environment may induce the net interactions between resonators to exhibit a certain directionality for non-trivial coupling phases $\phi_{1}$, $\phi_{2}$, $\phi_{3}$, $\theta$, and $\varphi$. For example, $|g_{3}e^{i\phi_{3}}-iJe^{i(\theta+\varphi)}|$ is not always equal to $|g_{3}e^{-i\phi_{3}}-iJe^{i(\theta+\varphi)}|$, which will result in the directional information transfer~\cite{PhysRevX.5.021025,PRL125023603}. We note that in microwave optomechanical circuits, although the common environment (e.g., substrate materials, microwave waveguides, or transmission lines) may not be directly regulated, the forms and strengths of the interactions between optomechanical resonators and the common environment can be adjusted in many ways~\cite{PhysRep.718.1,RMP85623}. Therefore, the interaction between the system and the common environment can be engineered. This means that the parameters $J$, $\theta$, and $\varphi$ can be tuned by engineering the interactions between resonators and the common environment. Based on different choices of parameters in Eq.~(\ref{Eq13}), we propose four special microwave optomechanical circuits, as illustrated in Fig.~\ref{Fig4}, and study their synchronization dynamics.

\begin{figure}[t]
    \centering
    \includegraphics[width=8.4cm]{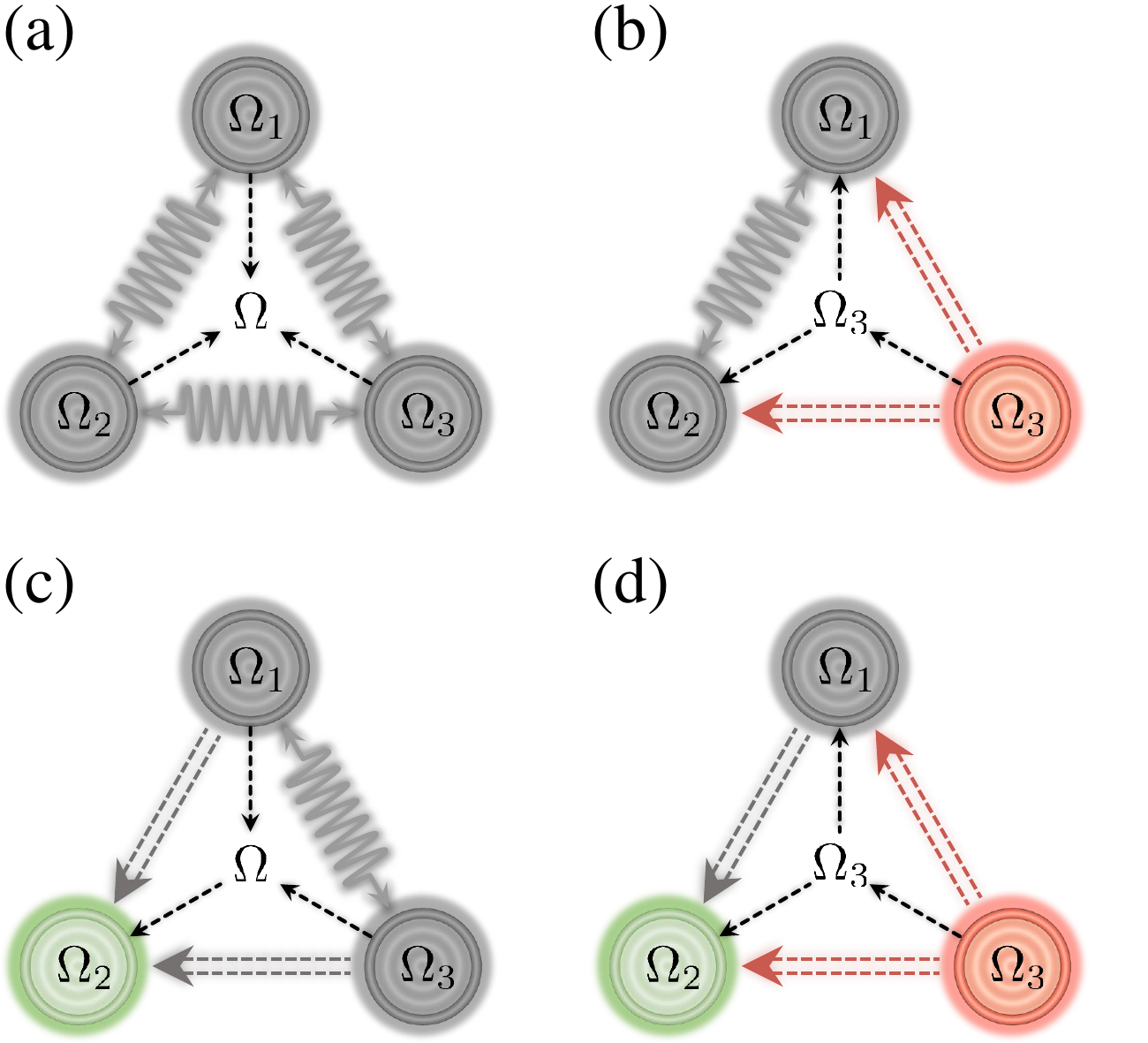}
    \caption{(a) Schematic diagram of a circuit in which there is a bidirectional interaction between any two resonators, with the interaction strength differing in each direction. (b) Schematic diagram of a circuit with both bidirectional and unidirectional interactions between resonators. We here assume that the red resonator with resonance frequency $\Omega_{3}$ is the ``output port", and the red dashed arrows represent the unidirectional (i.e., output) interactions between the ``output port'' resonator and the other resonators. (c) Schematic diagram of a circuit with both bidirectional and unidirectional interactions between the resonators. In this circuit, the green resonator with resonance frequency $\Omega_{2}$ is the ``input port", and the gray dashed arrows represent the unidirectional (i.e., input) interactions between the other resonators and the ``input port'' resonator. (d) Schematic diagram of a circuit in which only an unidirectional interaction exists between any two resonators, and the circuit contains both an ``output port'' resonator and an ``input port'' resonator.}
	\label{Fig4}
	\vspace{-0.3cm}
\end{figure}

\subsection{Synchronization of the circuit in Fig.~\ref{Fig4}(a)}

Figure~\ref{Fig4}(a) schematically illustrates a circuit described by the Hamiltonian in Eq.~(\ref{Eq13}), in which there is a bidirectional interaction between any two resonators, but the interaction strengths are different for two directions. That is, the coupling strengths of $\hat{a}_{1}^{\dag}\hat{a}_{2}$ and $\hat{a}_{2}^{\dag}\hat{a}_{1}$ in Eq.~(\ref{Eq13}) satisfy the condition
\begin{equation}
|g_{1}e^{i\phi_{1}}-iJe^{i\theta}|\neq|g_{1}e^{-i\phi_{1}}-iJe^{i\theta}|\neq0.
\label{Eq14}
\end{equation}
The coupling strengths of $\hat{a}_{2}^{\dag}\hat{a}_{3}$ and $\hat{a}_{3}^{\dag}\hat{a}_{2}$ in Eq.~(\ref{Eq13}) satisfy the condition
\begin{align}
|g_{2}e^{i\phi_{2}}-iJe^{i\varphi}|\neq|g_{2}e^{-i\phi_{2}}-iJe^{i\varphi}|\neq0.
\label{Eq15}
\end{align}
The coupling strengths of $\hat{a}_{3}^{\dag}\hat{a}_{1}$ and $\hat{a}_{1}^{\dag}\hat{a}_{3}$ in Eq.~(\ref{Eq13}) satisfy the condition
\begin{align}
|g_{3}e^{i\phi_{3}}-iJe^{i(\theta+\varphi)}|\neq|g_{3}e^{-i\phi_{3}}-iJe^{i(\theta+\varphi)}|\neq0.
\label{Eq16}
\end{align}

\begin{figure}[t]
	\centering
	\includegraphics[width=8.4cm]{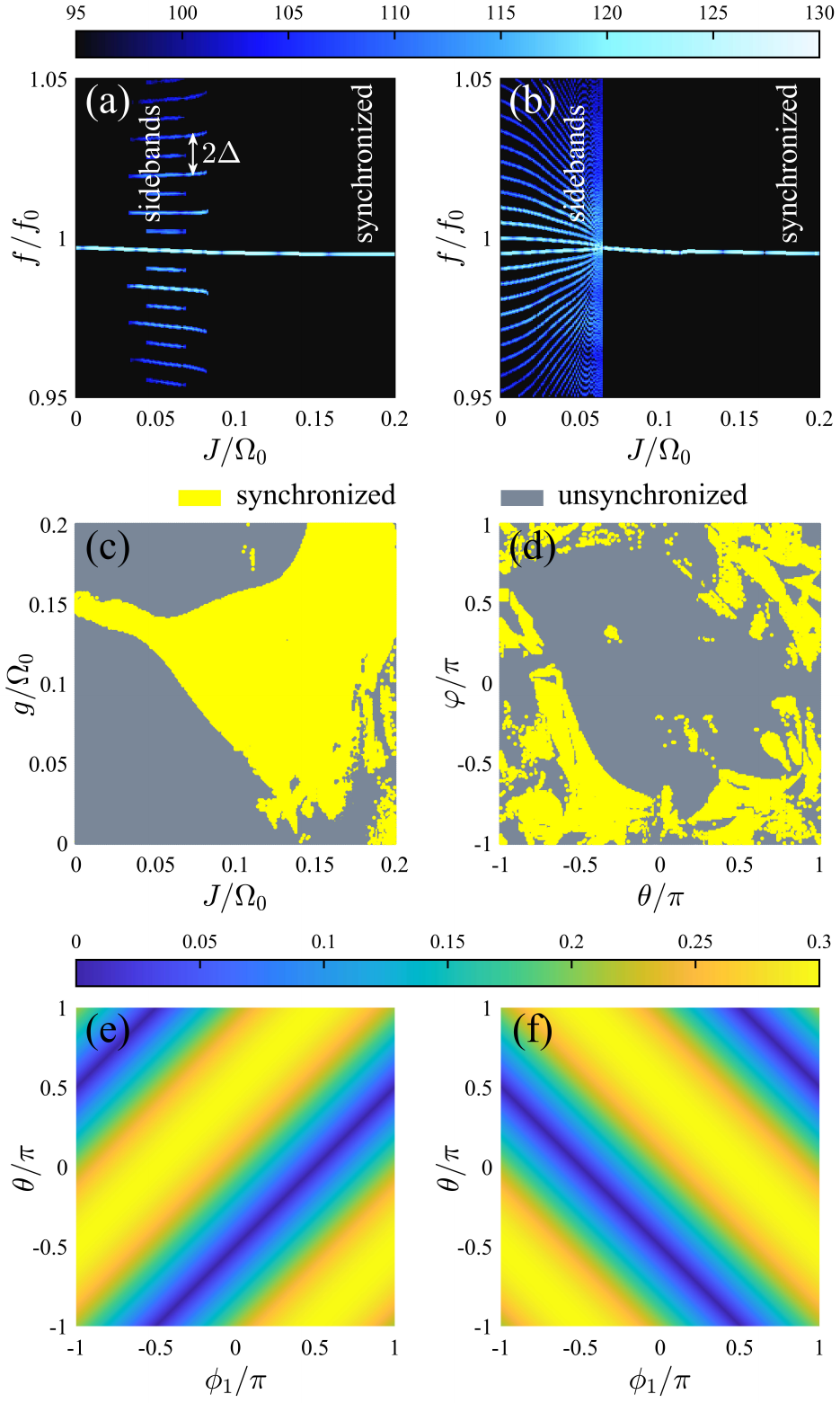}
	\caption{Frequency spectra of the circuit as a function of the dissipative coupling strength $J$, in (a), $g/\Omega_{0}=0.15$, $\theta=0.5\pi$, $\varphi=0.9\pi$, $\phi_{0}=0.8\pi$; in (b), $g/\Omega_{0}=0.15$, $\theta=0.5\pi$, $\phi_{0}=\pi$, $\varphi=-\;\!0.7\pi$. Phase diagrams of the system in the $(J,g)$ plane in (c) for $\phi_{0}=0.8\pi$, $\theta=0.5\pi$, $\varphi=-\;\!0.7\pi$, and in the $(\theta,\varphi)$ plane in (d) for $g/\Omega_{0}=J/\Omega_{0}=0.15$, $\phi_{0}=0.8\pi$, where the yellow and gray regions represent the frequency synchronization and unsynchronization regions, respectively. The value distributions of $|g_{1}e^{i\phi_{1}}-iJ e^{i\theta}|$ and $|g_{1}e^{-i\phi_{1}}-iJ e^{i\theta}|$ within the $(\phi_{1},\theta)$ plane are shown in (e) and (f) with $g/\Omega_{0}=J/\Omega_{0}=0.15$. In all figures, we assume $g_{1}=g_{2}=g_{3}=g$ and $\phi_{1}=\phi_{2}=\phi_{3}=\phi_{0}$, and other parameters are the same as those in Fig.~\ref{Fig2}.}
	\label{Fig5}
	\vspace{-0.3cm}
\end{figure}

In such a case, if the three resonators are synchronized, then the corresponding synchronized frequency $\Omega$ will be affected by the original frequency ($\Omega_{j}$, $j=1,2,3$) of each resonator. As shown in Figs.~\ref{Fig5}(a) and~\ref{Fig5}(b), regardless of the oscillation state of the circuit induced by the coherent coupling Hamiltonian $\hat{H}_{0}$, the common environment can effectively control the oscillation state. It should be noted that, different from the results shown in Fig.~\ref{Fig3}(a), more complex sideband peaks emerge because the interactions between resonators are nonreciprocal. Frequency spacing between neighboring sideband peaks in the spectrum can be $\Delta$, $2\Delta$, or even continuously variable. This means that multiple nonlinear effects are intertwined and the system dynamics become more complex. It is worth noting that the synchronized frequency is not fixed, but varies with the parameter $J$.

We find that the effects of the common environment ($J$, $\theta$, and $\varphi$) on the synchronization dynamics become more significant when the coherent interactions exist between resonators, as shown in Figs.~\ref{Fig5}(c) and~\ref{Fig5}(d). In this case, the common environment not only regulates the interaction strengths in Eq.~(\ref{Eq13}), but also regulates the nonreciprocity of the interactions in Eqs.~(\ref{Eq14})--(\ref{Eq16}). To more intuitively show the regulation of the common environment on the interactions between resonators, we here take $|g_{1}e^{i\phi_{1}}-iJe^{i\theta}|$ (for $\hat{a}_{1}^\dagger\hat{a}_{2}$) and $|g_{1}e^{-i\phi_{1}}-iJe^{i\theta}|$ (for $\hat{a}_{2}^\dagger\hat{a}_{1}$) in Eq.~(\ref{Eq14}) as examples to depict Figs.~\ref{Fig5}(e) and~\ref{Fig5}(f). The same results can be obtained from Eq.~(\ref{Eq15}) and Eq.~(\ref{Eq16}). By comparing Fig.~\ref{Fig5}(e) with Fig.~\ref{Fig5}(f), we conclude that the nonreciprocity of the interactions can be tuned continuously via the common environment. We also find that at some singular points, e.g., $\phi_{1}-\theta=0.5\pi$ in Fig.~\ref{Fig5}(e), the net interaction between resonators is unidirectional. One can easily verify that for $g_{1}=J$ and $\phi_{1}-\theta=0.5\pi$,
\begin{equation}
|g_{1}e^{i\phi_{1}}-iJe^{i\theta}|=0,\ \ |g_{1}e^{-i\phi_{1}}-iJe^{i\theta}|=2J|{\rm cos}\;\;\!\!\!\theta|.
\label{Eq17}
\end{equation}
Clearly, the synchronization dynamics of the circuit are regulated by the common environment.

\subsection{Synchronization of the circuit in Fig.~\ref{Fig4}(b)}

\begin{figure}[t]
	\centering
	\includegraphics[width=8.5cm]{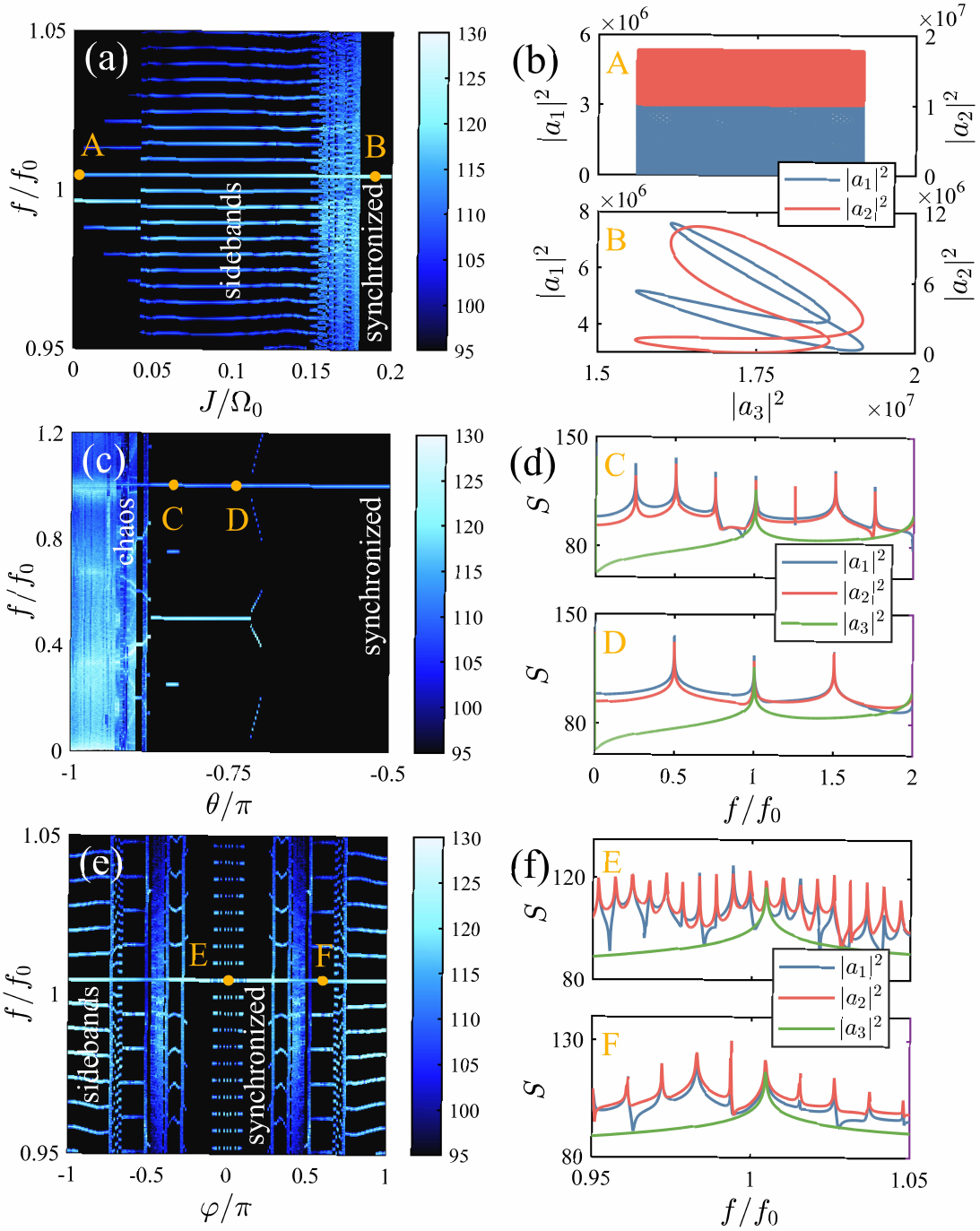}
	\caption{Frequency spectra of the circuit versus the parameters $J$ in (a), $\theta$ in (c), and $\varphi$ in (e), respectively. (b) Trajectories of $|a_{3}|^2$, $|a_{1}|^2$ (the blue curves) and $|a_{3}|^2$, $|a_{2}|^2$ (the red curves) for the representative points A (upper panel) and B (lower panel), in which A: $J=0$, B: $J/\Omega_{0}=0.19$. (d) Frequency spectra for the representative points C (upper panel) and D (lower panel), in which C: $\theta=-0.835\pi$, D: $\theta=-0.77\pi$. (f) Frequency spectra for the representative points E (upper panel) and F (lower panel), in which E: $\varphi=0$, F: $\varphi=0.58\pi$. The blue, red, and green curves represent $|a_{1}|^2$, $|a_{2}|^2$, and $|a_{3}|^2$, respectively. Parameters in (a) and (b) are $\varphi=-\;\!0.8\pi$, $\phi_{1}=0.7\pi$, $g_{1}/\Omega_{0}=0.18$, $\theta=0.5\pi$; in (c) and (d) are $g_{1}/\Omega_{0}=0.1$, $J/\Omega_{0}=0.2$, $\phi_{1}=0.6\pi$, $\varphi=0.3\pi$; in (e) and (f) are $g_{1}/\Omega_{0}=0.19$, $J/\Omega_{0}=0.17$, $\phi_{1}=0.3\pi$, $\theta=0$; other parameters are the same as those in Fig.~\ref{Fig2}.}
	\label{Fig6}
	\vspace{-0.3cm}
\end{figure}

As schematically shown in Eq.~(\ref{Eq13}), the interaction between any two resonators can be transformed from bidirectional to unidirectional by properly tuning the parameters. Based on this, we propose a special circuit in Fig.~\ref{Fig4}(b). Unlike the scenario presented in Fig.~\ref{Fig4}(a) where all resonators affect each other, in Fig.~\ref{Fig4}(b), the red resonator with resonance frequency $\Omega_{3}$ is not affected by other resonators, while it strongly affects the oscillation states of other resonators. The interaction Hamiltonian corresponding to the circuit in Fig.~\ref{Fig4}(b) can be written as
\begin{equation}
	\begin{split}
		\hat{H}=&\left(g_{1}e^{i\phi_{1}}-iJe^{i\theta}\right)\!\hat{a}_{1}^{\dag}\hat{a}_{2}-2iJ{\rm cos}(\theta+\varphi)\hat{a}_{1}^{\dag}\hat{a}_{3}\\+&\left(g_{1}e^{-i\phi_{1}}-iJe^{i\theta}\right)\!\hat{a}_{2}^{\dag}\hat{a}_{1}-2iJ{\rm cos}\;\;\!\!\!\varphi\;\;\!\!\!\hat{a}_{2}^{\dag}\hat{a}_{3},
		\label{Eq18}
	\end{split}
\end{equation}
which is derived from Eq.~(\ref{Eq13}) under the conditions $\phi_{3}=\varphi+\theta+0.5\pi$, $g_{3}=J$, $\phi_{2}=-\varphi-0.5\pi$, and $g_{2}=J$. An interesting conclusion can be inferred that if the three resonators achieve synchronized oscillations, the effective oscillation frequency can only be $\Omega_{3}$. However, due to the existence of the ``output port'' resonator represented by the red resonator, partial synchronization phenomena may occur in the circuit. To demonstrate these, we depict Fig.~\ref{Fig6}.

As shown in Fig.~\ref{Fig6}(a), when the dissipative coupling strength $J$ induced by the common environment is weak, i.e., $J\ll g_{1}$, the ``output port'' resonator with frequency $\Omega_{3}$ has a weak influence on the other two resonators. In this parameter region, partial synchronization of the resonators with frequencies $\Omega_{1}$ and $\Omega_{2}$ can occur, and the trajectories of $|a_{1}|^2$ and $|a_{2}|^2$ are nearly independent of $|a_{3}|^2$, which can be observed intuitively from the upper panel of Fig.~\ref{Fig6}(b). With the increase of $J$, the partial synchronization phenomenon (i.e., synchronization of the resonators with frequencies $\Omega_{1}$ and $\Omega_{2}$) will gradually disappear and be replaced by a series of sideband peaks related to $\Omega_{3}$. When $J$ exceeds a threshold, e.g., $J_{0}/\Omega_{0}=0.18$ in Fig.~\ref{Fig6}(a), then the three resonators in the circuit reach the synchronization region, i.e., all the resonators are synchronized to the frequency $\Omega_{3}$. The trajectories in the lower panel of Fig.~\ref{Fig6}(b) show the synchronization characteristics~\cite{PRL129063605}. Similar to the results in Fig.~\ref{Eq3}(b), after the resonators are synchronized to the frequency $\Omega_{3}$, if we further adjust the parameters, as shown in Fig.~\ref{Fig6}(c), multiple period-doubling bifurcations can be induced due to the nonlinearity of optomechanical interactions~\cite{PRL1140136012015}.

Our study shows that the state of the ``output port'' resonator is not affected by $J$, $\theta$, and $\varphi$, as demonstrated by the green curves in Fig.~\ref{Fig6}(d). The ``output port'' resonator oscillates consistently with the frequency $\Omega_{3}$. Thus, an interesting partial synchronization phenomenon of the resonators with frequencies $\Omega_{1}$ and $\Omega_{2}$ occurs in the circuit in Fig.~\ref{Eq4}(b). As shown by the blue (i.e., $|a_{1}|^2$) and red (i.e., $|a_{2}|^2$) curves in Fig.~\ref{Fig6}(d), the resonators other than the ``output port'' resonator are synchronized to $\Omega_{3}/N$, where $N$ represents a positive integer. For the upper panel of Fig.~\ref{Fig6}(d), $N=4$, and for the lower panel of Fig.~\ref{Fig6}(d), $N=2$. Moreover, we find that the distributions of sideband peaks in Figs.~\ref{Fig6}(a) and~\ref{Fig6}(e) always include a sideband peak at the frequency $\Omega_{3}$, which can be observed more clearly from Fig.~\ref{Fig6}(f). This indicates that the generation of sidebands is regulated by the ``output port'' resonator and the common environment.

\subsection{Synchronization of the circuit in Fig.~\ref{Fig4}(c)}

We now study another special circuit, as schematically shown in Fig.~\ref{Fig4}(c). Different from the red ``output port" resonator in Fig.~\ref{Fig4}(b), the green resonator in Fig.~\ref{Fig4}(c) with the frequency $\Omega_{2}$ can be considered as the ``input port" resonator. It has no influence on the other two resonators in the circuit, but its oscillation state is strongly influenced by the other resonators. The interaction Hamiltonian corresponding to the circuit in Fig.~\ref{Fig4}(c) can be written as
\begin{equation}
	\begin{split}
		\hat{H}=&\left[g_{3}e^{-i\phi_{3}}-iJe^{i(\theta+\varphi)}\right]\!\hat{a}_{1}^{\dag}\hat{a}_{3}-2iJ{\rm cos}\;\;\!\!\!\theta\;\;\!\!\!\hat{a}_{2}^{\dag}\hat{a}_{1}\\+&\left[g_{3}e^{i\phi_{3}}-iJe^{i(\theta+\varphi)}\right]\!\hat{a}_{3}^{\dag}\hat{a}_{1}-2iJ{\rm cos}\;\;\!\!\!\varphi\;\;\!\!\!\hat{a}_{2}^{\dag}\hat{a}_{3},
		\label{Eq19}
	\end{split}
\end{equation}
which is derived from Eq.~(\ref{Eq13}) under the conditions $\phi_{1}=\theta+0.5\pi$, $g_{1}=J$, $\phi_{2}=-\varphi-0.5\pi$, and $g_{2}=J$. We can infer that if the three optomechanical resonators realize synchronized oscillations, the synchronization frequency $\Omega$ will depend only on $\Omega_{1}$ and $\Omega_{3}$, i.e., independent of the ``input port'' resonator. Similar to the synchronization dynamics of the circuit in Fig.~\ref{Fig4}(b), a variety of partial synchronization phenomena also exist in this circuit. Unlike the ``output port'' resonator which is not affected by the other two resonators, the ``input port'' resonator is inevitably affected by the other two resonators. Thus, the synchronization dynamics of the circuit in Fig.~\ref{Fig4}(c) are more complicated.

\begin{figure}[t]
	\centering
	\includegraphics[width=8.2cm]{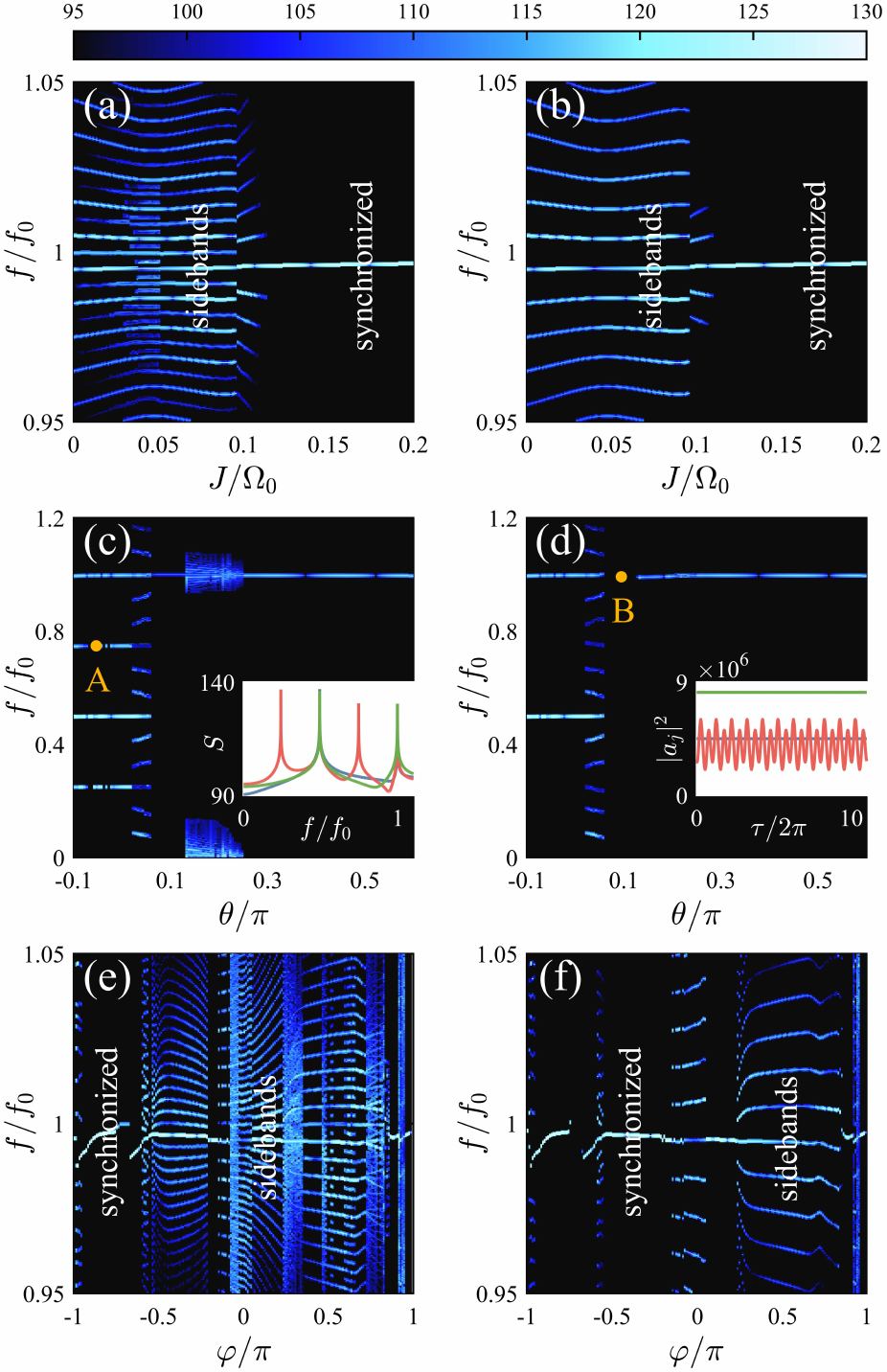}
	\caption{Frequency spectra of the circuit in Fig.~\ref{Fig4}(c) versus the parameters $J$ in (a), $\theta$ in (c), and $\varphi$ in (e), respectively. The inset in (c) shows the spectral structure for the representative point A, i.e., $\theta=-\;\!0.05\pi$, where the blue, red, and green curves correspond to $|a_{1}|^2$, $|a_{2}|^2$, and $|a_{3}|^2$, respectively. Frequency spectra of the circuit without the ``input port'' resonator versus the parameters $J$ in (b), $\theta$ in (d), and $\varphi$ in (f), respectively. The inset in (d) depicts the time-domain evolution of $|a_{j}|^2$ for the representative point B, i.e., $\theta=0.1\pi$, where the blue, red, and green curves represent $|a_{1}|^2$, $|a_{2}|^2$, and $|a_{3}|^2$, respectively. Parameters in (a) and (b) are $\varphi=-\;\!0.7\pi$, $\phi_{3}=0.7\pi$, $g_{3}/\Omega_{0}=0.2$, $\theta=0.2\pi$; in (c) and (d) are $g_{3}/\Omega_{0}=J/\Omega_{0}=0.16$, $\phi_{3}=\varphi=0.2\pi$; in (e) and (f) are $\phi_{3}=0.2\pi$, $\theta=\pi$, $g_{3}/\Omega_{0}=J/\Omega_{0}=0.16$; other parameters are the same as those in Fig.~\ref{Fig2}.}
	\label{Fig7}
	\vspace{-0.3cm}
\end{figure}

Figure~\ref{Fig7} illustrates the numerical simulations of the synchronization dynamics corresponding to the circuit shown in Fig.~\ref{Fig4}(c) for different parameter regions. From Figs.~\ref{Fig7}(a) and~\ref{Fig7}(b), we find that the two resonators with frequencies $\Omega_{1}$ and $\Omega_{3}$ can be synchronized to the frequency $\Omega$ ($\Omega\neq\Omega_{j}$ with $j=1,2,3$) with the increase of dissipative coupling strength $J$ for given parameters $\theta$, $\phi_{3}$, and $\varphi$. If $J$ is further increased, then the ``input port'' resonator will also be synchronized to the frequency $\Omega$. In the sideband region, the ``input port'' resonator may generate a series of sidebands which are different from those of other resonators. Thus, the structure of frequency spectra becomes very complicated. Moreover, as shown in Figs.~\ref{Eq7}(c) and~\ref{Eq7}(d), when the other two resonators are synchronized to the frequency $\Omega$, the state of ``input port'' resonator is still undetermined. It may be in the sideband region or synchronized to $\Omega/N$, where $N$ represents a positive integer.

The inset in Fig.~\ref{Eq7}(c) corresponds to the representative point A, which shows the case $N=2$. Here, it should be noted that for the circuits with ``input port'' resonator, period-doubling bifurcations will be more complicated. The inset in Fig.~\ref{Eq7}(d) corresponds to the representative point B, which depicts the phenomenon of ``oscillation death''. Figures.~\ref{Eq7}(e) and~\ref{Eq7}(f) illustrate the distributions of sideband peaks for the ``input port'' resonator and the other resonators. It is obvious that the sideband peaks of the ``input port'' resonator contain the sidebands of all other resonators, and the effective distance between peaks varies continuously as the parameters change.

\begin{figure}[t]
	\centering
	\includegraphics[width=8.4cm]{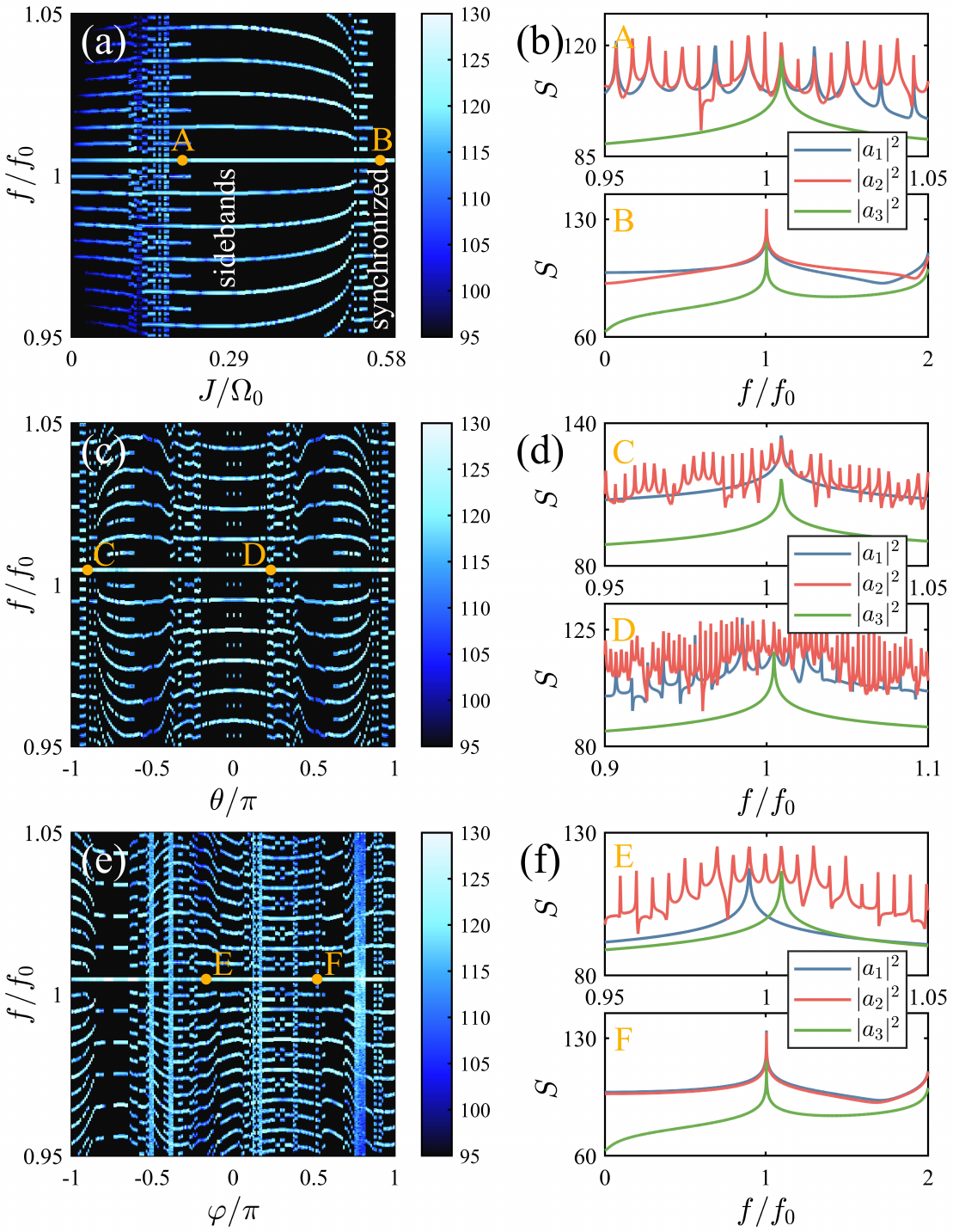}
	\caption{Frequency spectra of the circuit in Fig.~\ref{Fig4}(d) versus the parameters $J$ in (a), $\theta$ in (c), and $\varphi$ in (e), respectively. (b) Frequency spectra for the representative points A (upper panel) and B (lower panel), in which A: $J/\Omega_{0}=0.2$, B: $J/\Omega_{0}=0.57$. (d) Frequency spectra for the representative points C (upper panel) and D (lower panel), in which C: $\theta=-\;\!0.95\pi$, D: $\theta=0.24\pi$. (f) Frequency spectra for the representative points E (upper panel) and F (lower panel), in which E: $\varphi=-\;\!0.2\pi$, F: $\varphi=0.5\pi$. The blue, red, and green curves represent $|a_{1}|^2$, $|a_{2}|^2$, and $|a_{3}|^2$, respectively. Parameters in (a) and (b) are $\varphi=\pi$, $\theta=0.9\pi$; in (c) and (d) are $\varphi=\pi$, $J/\Omega_{0}=0.56$; in (e) and (f) are $J/\Omega_{0}=0.56$, $\theta=0.7\pi$; other parameters are the same as those in Fig.~\ref{Fig2}.}
	\label{Fig8}
	\vspace{-0.3cm}
\end{figure}

\subsection{Synchronization of the circuit in Fig.~\ref{Fig4}(d)}

We now study another type of circuit, as schematically shown in Fig.~\ref{Fig4}(d). This circuit contains an ``output port'' resonator (marked in red) and an ``input port'' resonator (marked in green). In such a circuit, the interaction between any two resonators is unidirectional, and the interaction Hamiltonian corresponding to the circuit in Fig.~\ref{Fig4}(d) is written as
\begin{align}
  \hat{H}=-\:2iJ\!\left[{\rm cos}(\theta+\varphi)\hat{a}_{1}^{\dag}\hat{a}_{3}+{\rm cos}\;\;\!\!\!\theta\;\;\!\!\!\hat{a}_{2}^{\dag}\hat{a}_{1}+{\rm cos}\;\;\!\!\!\varphi\;\;\!\!\!\hat{a}_{2}^{\dag}\hat{a}_{3}\right]\!\;\!\!,
  \label{Eq20}
\end{align}
which is derived from Eq.~(\ref{Eq13}) under the conditions $\phi_{1}=\theta+0.5\pi$, $g_{1}=J$, $\phi_{2}=-\varphi-0.5\pi$, $g_{2}=J$, $\phi_{3}=\varphi+\theta+0.5\pi$, and $g_{3}=J$. Due to the existence of the ``output port'' resonator, all resonators can only be synchronized to the frequency $\Omega_{3}$ of the ``output port'' resonator.

Synchronization dynamics of this circuit are presented in Fig.~\ref{Fig8}. Due to the unidirectional interactions between resonators, the interaction strength $J$ required to realize synchronization is stronger than that of the other circuits shown in Fig.~\ref{Fig4}. As shown in Figs.~\ref{Fig8}(a),~\ref{Fig8}(c), and~\ref{Fig8}(e), the parameter regions corresponding to the synchronization regions are very narrow. This indicates that the conditions for achieving synchronization in the circuit in Fig.~\ref{Fig4}(d) are more stringent than those for the other circuits in Fig.~\ref{Fig4}. Moreover, we find that the oscillation states of the resonators in Fig.~\ref{Fig4}(d) can be completely different, e.g., the representative point A in Fig.~\ref{Fig8}(b), which is different from the cases discussed previously. When the coupling strength is strong enough, e.g., $J/\Omega_{0}=0.57$, the three resonators can eventually be synchronized to the frequency $\Omega_{3}$, as shown in the lower panel of Fig.~\ref{Fig8}(b). By adjusting the parameters $J$, $\theta$, and $\varphi$, i.e., engineering the interactions between microwave optomechanical resonators and the common environment, we can also observe a series of partial synchronization phenomena in this circuit. For example, at the representative point C in Fig.~\ref{Fig8}(d), the resonators with frequencies $\Omega_{1}$ and $\Omega_{3}$ are synchronized to the frequency $\Omega_{3}$, while the oscillation state of the ``input port" resonator (i.e., the green resonator with frequency $\Omega_{2}$) exhibits a series of complex sideband peaks.

It should be emphasized here that the ``input port'' resonator contains the oscillation information from the other resonators. Thus, the dynamics of the ``input port'' resonator is highly sensitive to the variations of parameters. We find that the sideband peaks on the spectra of other resonators will appear at the corresponding positions on the spectrum of the ``input port'' resonator, e.g., the representative points A in Fig.~\ref{Fig8}(b) and D in Fig.~\ref{Fig8}(d). It is worth noting that the circuit shown in Fig.~\ref{Fig4}(d) can be further opened into a series of non-closed chains. If we decouple the resonators with frequencies $\Omega_{1}$ and $\Omega_{3}$, i.e., ${\rm cos}(\theta+\varphi)=0$, as shown in the representative point E in Fig.~\ref{Fig8}(e), then synchronization of the resonators will never occur ($\Omega_{1}\neq\Omega_{2}\neq\Omega_{3}$). This phenomenon indicates that the synchronization in the circuit is suppressed by the common environmental effects. The dynamics of the ``input port'' resonator will be controlled by two independent nonidentical resonators, and the final oscillation state is indeterminate (there is a competing process). Interestingly, as shown in the representative point F in Fig.~\ref{Fig8}(f), i.e., ${\rm cos}\varphi=0$, in which the resonators with frequencies $\Omega_{2}$ and $\Omega_{3}$ are decoupled, synchronized oscillations of the resonators can still be achieved under certain conditions. This demonstrates that by exploiting the common environment, we can achieve flexible control of the synchronization in microwave optomechanical circuits. There are many other variations of the circuit in Fig.~\ref{Fig4}(d), and the synchronization dynamics in these variations deserve to be further studied.

\section{Discussion and conclusion \label{sec5}}

In summary, we have introduced the common environment as a new approach to study the synchronization dynamics of microwave optomechanical circuits. By eliminating the environmental modes, the effects of common environment on the dynamics can be described using an equivalent non-Hermitian Hamiltonian. We demonstrate that the frequency synchronization of nonidentical microwave optomechanical resonators can be achieved via a common environment. Moreover, we show that the synchronization dynamics of conventional microwave circuits (i.e., only the Hermitian interactions exist between resonators) can be significantly regulated by the non-Hermitian interactions mediated by the common environment. By engineering the interactions between resonators and the common environment, the net interactions between resonators can be transformed from bidirectional to unidirectional. Therefore, the synchronization phenomena in the circuits are greatly enriched. We propose and study four special circuits with both Hermitian and non-Hermitian interactions between resonators, and observe many anomalous synchronization behaviors in these circuits.

Microwave optomechanical circuits based on superconducting circuits have mature fabrication processes, great scalability, and superior controllability~\cite{RMP861391,PhysRep.718.1,RMP93025005}. Thus, they can serve as potential platforms for synchronization research. Common environments are ubiquitous in solid-state many-body systems~\cite{Many-body book}, and in microwave circuits, they can be substrate materials, microwave transmission lines, cavities, or waveguides, which connect two or more subsystems~\cite{RMP85623}. Moreover, the form and strength of the interaction between the common environment and the resonators are engineerable~\cite{PhysRevX.5.021025,NC86042017,En12019}, and the interaction strength between the environment and the subsystems can reach the superstrong or even ultrastrong coupling regime~\cite{PhysRevLett.127.250402,RMP91025005}. The tunability of coherent interaction between resonators is important in our proposal, but this requirement is not difficult for superconducting circuits~\cite{PhysRep.718.1}. In recent years, various high-quality couplers have been proposed~\cite{PhysRevLett.130.030603,PhysRevB.87.134504,PRXquantum22021,naturephysics2017.13}, and the coupling strength between resonators can exceed $400\;\rm MHz$~\cite{nature6662022,RMP861391,naturephysics2017.13.465}. Therefore, the parameter region for the synchronization phenomena observed in this paper is realistic under current experimental conditions. We hope that this work will inspire researchers to apply the common environment to synchronization research. Our study may open up new perspectives for the construction of microwave synchronization networks.

\begin{acknowledgments}
This work was supported by the National Natural Science Foundation of China under Grant No.~12374483 and No.~92365209.
\end{acknowledgments}

\appendix
\section{non-Hermitian dissipative coupling induced by the common environment \label{Appendix1}}

In this appendix, we provide the detailed procedure for deriving the non-Hermitian dissipative coupling between resonators induced by the common environment. We define the microwave optomechanical circuit as the system and the transmission line interacting with each resonator as the common environment. Then the Hamiltonian $\hat{H}_{\rm OM}$ of the system reads (henceforth, we set $\hbar=1$)
\begin{equation}
	\hat{H}_{\rm OM}=\sum_{j=1}^{3}\omega_{j}\hat{a}_{j}^{\dagger}\hat{a}_{j}+\Omega_{j}\hat{b}_{j}^{\dagger}\hat{b}_{j}-G_{j}\hat{a}_{j}^{\dag}\hat{a}_{j}(\hat{b}_{j}^{\dagger}+\hat{b}_{j}),
	\label{A1}
\end{equation}
where $\hat{a}_{j}$ and $\hat{b}_{j}$ denote the annihilation operators of the microwave mode with frequency $\omega_{j}$ and the mechanical mode with frequency $\Omega_{j}$ supported by the $j$-th resonator, respectively, and $G_{j}$ is the single-photon optomechanical coupling strength of the $j$-th resonator. The Hamiltonian $\hat{H}_{\rm E}$ of the common environment reads
\begin{equation}
	\hat{H}_{\rm E}=\int_{0}^{\infty}\!d\omega\;\;\!\!\!\omega\!\left[\hat{c}_{\rm L}^{\dag}(\omega)\hat{c}_{\rm L}(\omega)+\hat{c}_{\rm R}^{\dag}(\omega)\hat{c}_{\rm R}(\omega)\right]\!\;\!\!,
	\label{A2}
\end{equation}
where $\hat{c}_{\rm L(R)}(\omega)$ and $\hat{c}_{\rm L(R)}^{\dag}(\omega)$ denote the annihilation and creation operators of the left (right) moving environment mode at frequency $\omega$. For simplicity, we assume that the coupling between resonators and the transmission line is only through the capacitors~\cite{PhysRep.718.1}. Then we can obtain the interaction Hamiltonian $\hat{H}_{\rm I}$ between the system and the common environment as~\cite{PhysRep.718.1,PRA88043806}
\begin{equation}
   \begin{split}
	\hat{H}_{\rm I}=&\sum_{j=1}^{3}\int_{0}^{\infty}\!d\omega J_{j}
	\sqrt{\omega}\;\;\!\!\!\hat{a}_{j}^{\dagger}\!\left[\hat{c}_{\rm L}(\omega)e^{-i\omega x_{j}/v}+\hat{c}_{\rm R}(\omega)e^{i\omega x_{j}/v}\right] \\
	+&\sum_{j=1}^{3}\int_{0}^{\infty}\!d\omega J_{j}
	\sqrt{\omega}\;\;\!\!\!\hat{a}_{j}\!\left[\hat{c}_{\rm L}^{\dagger}(\omega)e^{i\omega x_{j}/v}+\hat{c}_{\rm R}^{\dagger}(\omega)e^{-i\omega x_{j}/v}\right]\!\;\!\!.
	\label{A3}
	\end{split}
\end{equation}
$x_{j}$ is the position of the $j$-th resonator, $J_{j}$ is the coupling strength between the $j$-th resonator and the transmission line, and $v$ denotes the group velocity of environmental modes. It is worth noting that the size $l$ of the microwave optomechanical resonator is very small ($l\sim10$ $\mu$m~\cite{Nature2011204471}). In experiments~\cite{RMP85623}, the length of the transmission line is $L\sim10$ cm. Thus, we can assume that the resonator is coupled to a certain position $x_{j}$ on the transmission line, and the capacitance of the resonator can be regarded as a small perturbation on the entire transmission line.

For generality, we define an arbitrary system operator $\hat{o}$, and the Heisenberg equation for $\hat{o}$ is
\begin{widetext}
\begin{align}
    i\dfrac{d\hat{o}(t)}{dt}&=\left[\hat{o}(t),\hat{H}_{\rm OM}(t)\right]+\sum_{j=1}^{3}\int_{0}^{\infty}\!d\omega J_{j}
    \sqrt{\omega}\Big[\hat{o}(t),\hat{a}_{j}^{\dag}(t)\Big]\!\!\;\!\left(\hat{c}_{\rm L}(\omega,t)e^{-i\omega  x_{j}/v}+\hat{c}_{\rm R}(\omega,t)e^{i\omega x_{j}/v}\right) \nonumber \\
    &+\sum_{j=1}^{3}\int_{0}^{\infty}\!d\omega J_{j}
    \sqrt{\omega}\!\left(\hat{c}_{\rm L}^{\dag}(\omega)e^{i\omega x_{j}/v}+\hat{c}_{\rm R}^{\dag}(\omega)e^{-i\omega x_{j}/v}\right)\!\!\;\!\Big[\hat{o}(t),\hat{a}_{j}(t)\Big]\;\!\!.
	\label{A4}
\end{align}
\end{widetext}
The Heisenberg equations for $\hat{c}_{\rm L}(\omega,t)$ and $\hat{c}_{\rm R}(\omega,t)$ can be written as
\begin{equation}
	\begin{split}
		\dfrac{d\hat{c}_{\rm L}(\omega,t)}{dt}=&-i\omega\hat{c}_{\rm L}(\omega,t)-i\sum_{j=1}^{3} J_{j}\sqrt{\omega}\;\;\!\!\!\hat{a}_{j}(t)e^{i\omega x_{j}/v},\\
		\dfrac{d\hat{c}_{\rm R}(\omega,t)}{dt}=&-i\omega\hat{c}_{\rm R}(\omega,t)-i\sum_{j=1}^{3} J_{j}\sqrt{\omega}\;\;\!\!\!\hat{a}_{j}(t)e^{-i\omega x_{j}/v}.
		\label{A5}
	\end{split}
\end{equation}
Further processing Eq.~(\ref{A5}) by formal integration gives
\begin{align}
	\hat{c}_{\rm L}(\omega,t)&=\hat{c}_{\rm L}(\omega,0)e^{-i\omega t} \nonumber \\
	&-i\sum_{j=1}^{3} J_{j}\sqrt{\omega}\!\int_{0}^{t}\!\hat{a}_{j}(t')e^{-i\omega(t-t'-x_{j}/v)}dt', \nonumber \\
	\hat{c}_{\rm R}(\omega,t)&=\hat{c}_{\rm R}(\omega,0)e^{-i\omega t} \nonumber \\
	&-i\sum_{j=1}^{3} J_{j}\sqrt{\omega}\!\int_{0}^{t}\!\hat{a}_{j}(t')e^{-i\omega(t-t'+x_{j}/v)}dt'. \label{A6}
\end{align}
For simplicity, we separate the slowly varying and rapidly oscillatory contributions of $\hat{a}_{j}(t)$, i.e., $\hat{a}_{j}(t)=\tilde{a}_{j}(t)e^{-i\omega_{0}t}$, which is equivalent to using a rotating reference frame at frequency $\omega_{0}$. Furthermore, it should be emphasized that in the constructed microwave optomechanical circuit, the resonance frequencies of the microwave and mechanical modes of the optomechanical resonators are only slightly detuned, that is, $(\omega_{j}-\omega_{0})\ll\omega_{0}$ and $(\Omega_{j}-\Omega_{0})\ll\Omega_{0}$ for $j=1,2,3$. By substituting the results in Eq.~(\ref{A6}) into Eq.~(\ref{A4}), we can rewrite Eq.~(\ref{A4}) as
\begin{widetext}
	\begin{align}
		i\dfrac{d\hat{o}(t)}{dt}&=\left[\hat{o}(t),\hat{H}_{\rm OM}(t)\right]+\sum_{j=1}^{3}\int_{0}^{\infty}\!d\omega J_{j}
		\sqrt{\omega}\Big[\hat{o}(t),\hat{a}_{j}^{\dag}(t)\Big]\!\;\!\!\left(\hat{c}_{\rm L}(\omega,0)e^{-i\omega t}e^{-i\omega  x_{j}/v}+\hat{c}_{\rm R}(\omega,0)e^{-i\omega t}e^{i\omega x_{j}/v}\right) \nonumber \\
		&+\sum_{j=1}^{3}\int_{0}^{\infty}\!d\omega J_{j}
		\sqrt{\omega}\!\left(\hat{c}_{\rm L}^{\dagger}(\omega,0)e^{i\omega t}e^{i\omega x_{j}/v}+\hat{c}_{\rm R}^{\dagger}(\omega,0)e^{i\omega t}e^{-i\omega x_{j}/v}\right)\!\!\;\!\Big[\hat{o}(t),\hat{a}_{j}(t)\Big] \nonumber \\
		&-i\sum_{j,k=1}^{3}\int_{0}^{\infty}\!d\omega J_{j}J_{k}\;\;\!\!\!
		\omega\Big[\hat{o}(t),\tilde{a}_{j}^{\dag}(t)\Big]\!\int_{0}^{t}\!\tilde{a}_{k}(t')e^{-i(\omega-\omega_{0})(t-t')}\Big(e^{i\omega (x_{k}-\;\;\!\!\!x_{j})/v}+e^{i\omega (x_{j}-\;\;\!\!\!x_{k})/v}\Big)dt' \nonumber \\
		&+i\sum_{j,k=1}^{3}\int_{0}^{\infty}\!d\omega J_{j}J_{k}\;\;\!\!\!\omega\!\int_{0}^{t}\!\tilde{a}_{k}^{\dagger}(t')e^{i(\omega-\omega_{0})(t-t')}\Big(e^{i\omega (x_{j}-\;\;\!\!\!x_{k})/v}+e^{i\omega (x_{k}-\;\;\!\!\!x_{j})/v}\Big)\!\;\!\Big[\hat{o}(t),\tilde{a}_{j}(t)\Big]dt'.
		\label{A7}
	\end{align}
\end{widetext}
We further assume that the system and the environment are initially uncorrelated, and the environment is in the vacuum state, i.e., $|\rm vac\rangle$. Therefore, it is straightforward to obtain
\begin{equation}
\langle\hat{c}_{\rm L}(\omega,0)\rangle=\langle\hat{c}_{\rm L}^{\dagger}(\omega,0)\rangle=\langle\hat{c}_{\rm R}(\omega,0)\rangle=\langle\hat{c}_{\rm R}^{\dagger}(\omega,0)\rangle.
\end{equation}
In the following discussion, we neglect the environmental modes in Eq.~(\ref{A7}). We here assume that the time scales of the evolution of system operators are much longer than the correlation time of the common environment~\cite{PRA042116,PRA88043806} (under the Born-Markov approximation). Then Eq.~(\ref{A7}) can be simplified to
\begin{widetext}
	\begin{align}
		i\dfrac{d\hat{o}(t)}{dt}&=\left[\hat{o}(t),\hat{H}_{\rm OM}(t)\right]-i\sum_{j,k=1}^{3}2\pi J_{j}J_{k}\;\;\!\!\!\omega_{0}\Big[\hat{o}(t),\tilde{a}_{j}^{\dag}(t)\Big]\!\;\!\Big(\tilde{a}_{k}(t-t_{kj})e^{i\omega_{0}t_{kj}}\Theta(t_{kj})+\tilde{a}_{k}(t+t_{kj})e^{-i\omega_{0}t_{kj}}\Theta(-t_{kj})\Big) \nonumber \\
		&+i\sum_{j,k=1}^{3}2\pi J_{j}J_{k}\;\;\!\!\!\omega_{0}\Big(\tilde{a}_{k}^{\dagger}(t-t_{kj})e^{-i\omega_{0}t_{kj}}\Theta(t_{kj})+\tilde{a}_{k}^{\dagger}(t+t_{kj})e^{i\omega_{0}t_{kj}}\Theta(-t_{kj})\Big)\!\;\!\Big[\hat{o}(t),\tilde{a}_{j}(t)\Big]\nonumber \\
		&-i\sum_{j=1}^{3}2\pi J_{j}^{2}\;\;\!\!\!\omega_{0}\Big[\hat{o}(t),\tilde{a}_{j}^{\dag}(t)\Big]\tilde{a}_{k}(t)+i\sum_{k=1}^{3}2\pi J_{k}^{2}\;\;\!\!\!\omega_{0}\tilde{a}_{k}^{\dagger}(t)\Big[\hat{o}(t),\tilde{a}_{j}(t)\Big]\!\;\!,
		\label{A9}
	\end{align}
\end{widetext}
in which $t_{kj}=(x_{k}-x_{j})/v$, and the function $\Theta(x)$ is defined as $\Theta(x)=1$ for $x>0$ and $\Theta(x)=0$ for $x\leq0$. We neglect the retardation effect caused by the propagation velocity $v$ of photons, i.e., $\tilde{a}_{k}(t\pm t_{kj})\approx\tilde{a}_{k}(t)$. This approximation is reasonable if the time scales on which system operators evolve are much slower than the time scale $t_{kj}$ for photons propagating in the common environment, e.g., transmission lines and waveguides in microwave circuits. If we further assume $x_{3}>x_{2}>x_{1}$, and define the phase parameters $\theta$ and $\varphi$ as
\begin{equation}
	\theta=\omega_{0}|x_{2}-x_{1}|/v, \quad \varphi=\omega_{0}|x_{3}-x_{2}|/v,
	\label{A10}
\end{equation}
then we can obtain the non-Hermitian equation of motion obeyed by $\hat{o}(t)$
\begin{widetext}
	\begin{align}
		i\dfrac{d\hat{o}(t)}{dt}&=\left[\hat{o}(t),\hat{H}_{\rm OM}(t)\right]-i\sum_{j=1}^{3}2\pi J_{j}^{2}\;\;\!\!\!\omega_{0}\Big[\hat{o}(t),\hat{a}_{j}^{\dag}(t)\Big]\hat{a}_{j}(t)+i\sum_{k=1}^{3}2\pi J_{k}^{2}\;\;\!\!\!\omega_{0}\hat{a}_{k}^{\dagger}(t)\Big[\hat{o}(t),\hat{a}_{k}(t)\Big] \nonumber \\
       &+i2\pi J_{1}J_{2}\;\;\!\!\!\omega_{0}\hat{a}_{2}^{\dagger}(t)e^{-i\theta}\Big[\hat{o}(t),\hat{a}_{1}(t)\Big]+i2\pi J_{1}J_{2}\;\;\!\!\!\omega_{0}\hat{a}_{1}^{\dagger}(t)e^{-i\theta}\Big[\hat{o}(t),\hat{a}_{2}(t)\Big]+i2\pi J_{1}J_{3}\;\;\!\!\!\omega_{0}\hat{a}_{3}^{\dagger}(t)e^{-i(\theta+\varphi)}\Big[\hat{o}(t),\hat{a}_{1}(t)\Big] \nonumber \\
       &+i2\pi J_{1}J_{3}\;\;\!\!\!\omega_{0}\hat{a}_{1}^{\dagger}(t)e^{-i(\theta+\varphi)}\Big[\hat{o}(t),\hat{a}_{3}(t)\Big]+i2\pi J_{2}J_{3}\;\;\!\!\!\omega_{0}\hat{a}_{3}^{\dagger}(t)e^{-i\varphi}\Big[\hat{o}(t),\hat{a}_{2}(t)\Big]+i2\pi J_{2}J_{3}\;\;\!\!\!\omega_{0}\hat{a}_{2}^{\dagger}(t)e^{-i\varphi}\Big[\hat{o}(t),\hat{a}_{3}(t)\Big]
       \nonumber \\
       &-i2\pi J_{1}J_{2}\;\;\!\!\!\omega_{0}\Big[\hat{o}(t),\hat{a}_{1}^{\dagger}(t)\Big]\hat{a}_{2}(t)e^{i\theta}-i2\pi J_{1}J_{2}\;\;\!\!\!\omega_{0}\Big[\hat{o}(t),\hat{a}_{2}^{\dagger}(t)\Big]\hat{a}_{1}(t)e^{i\theta}-i2\pi J_{1}J_{3}\;\;\!\!\!\omega_{0}\Big[\hat{o}(t),\hat{a}_{1}^{\dagger}(t)\Big]\hat{a}_{3}(t)e^{i(\theta+\varphi)} \nonumber \\
       &-i2\pi J_{1}J_{3}\;\;\!\!\!\omega_{0}\Big[\hat{o}(t),\hat{a}_{3}^{\dagger}(t)\Big]\hat{a}_{1}(t)e^{i(\theta+\varphi)}-i2\pi J_{2}J_{3}\;\;\!\!\!\omega_{0}\Big[\hat{o}(t),\hat{a}_{2}^{\dagger}(t)\Big]\hat{a}_{3}(t)e^{i\varphi}-i2\pi J_{2}J_{3}\;\;\!\!\!\omega_{0}\Big[\hat{o}(t),\hat{a}_{3}^{\dagger}(t)\Big]\hat{a}_{2}(t)e^{i\varphi}.
		\label{A11}
	\end{align}
\end{widetext}
We note that this equation holds for all system operators. Therefore, by utilizing the equivalence relation ${\rm Tr}(\dot{a}\rho)={\rm Tr}(\dot{\rho}a)$, we can move the time dependence in the system operator $\hat{o}(t)$ to the system density operator $\hat{\rho}(t)$.

Finally, the reduced master equation (after eliminating the common environment) for the evolution of the system density operator $\hat{\rho}(t)$ can be obtained as
\begin{widetext}
	\begin{align}
		i\dfrac{d\hat{\rho}(t)}{dt}&=\left[\hat{H}_{\rm OM},\hat{\rho}(t)\right]-i\sum_{j=1}^{3}2\pi J_{j}^{2}\;\;\!\!\!\omega_{0}\Big[\hat{a}_{j}^{\dag}\hat{a}_{j}\hat{\rho}(t)-2\hat{a}_{j}\hat{\rho}(t)\hat{a}_{j}^{\dagger}+\hat{\rho}(t)\hat{a}_{j}^{\dag}\hat{a}_{j}\Big] \nonumber \\
		&-i2\pi J_{1}J_{2}\;\;\!\!\!\omega_{0}\Big[(\hat{a}_{2}^{\dagger}\hat{a}_{1}e^{i\theta}+\hat{a}_{1}^{\dagger}\hat{a}_{2}e^{i\theta})\hat{\rho}(t)+\hat{\rho}(t)(\hat{a}_{2}^{\dagger}\hat{a}_{1}e^{-i\theta}+\hat{a}_{1}^{\dagger}\hat{a}_{2}e^{-i\theta})-2{\;\;\!\!\!\rm cos\;\;\!\!\!}\theta\hat{a}_{2}\hat{\rho}(t)\hat{a}_{1}^{\dagger}-2{\;\;\!\!\!\rm cos\;\;\!\!\!}\theta\hat{a}_{1}\hat{\rho}(t)\hat{a}_{2}^{\dagger}\Big] \nonumber \\
		&-i2\pi J_{2}J_{3}\;\;\!\!\!\omega_{0}\Big[(\hat{a}_{3}^{\dagger}\hat{a}_{2}e^{i\varphi}+\hat{a}_{2}^{\dagger}\hat{a}_{3}e^{i\varphi})\hat{\rho}(t)+\hat{\rho}(t)(\hat{a}_{3}^{\dagger}\hat{a}_{2}e^{-i\varphi}+\hat{a}_{2}^{\dagger}\hat{a}_{3}e^{-i\varphi})-2{\;\;\!\!\!\rm cos\;\;\!\!\!}\varphi\hat{a}_{3}\hat{\rho}(t)\hat{a}_{2}^{\dagger}-2{\;\;\!\!\!\rm cos\;\;\!\!\!}\varphi\hat{a}_{2}\hat{\rho}(t)\hat{a}_{3}^{\dagger}\Big] \nonumber \\
		&-i2\pi J_{1}J_{3}\;\;\!\!\!\omega_{0}\Big\{\!\left[\hat{a}_{3}^{\dagger}\hat{a}_{1}e^{i(\theta+\varphi)}+\hat{a}_{1}^{\dagger}\hat{a}_{3}e^{i(\theta+\varphi)}\right]\!\hat{\rho}(t)+\hat{\rho}(t)\!\left[\hat{a}_{3}^{\dagger}\hat{a}_{1}e^{-i(\theta+\varphi)}+\hat{a}_{1}^{\dagger}\hat{a}_{3}e^{-i(\theta+\varphi)}\right] \nonumber\\
		&-2{\;\;\!\!\!\rm cos}(\theta+\varphi)\hat{a}_{3}\hat{\rho}(t)\hat{a}_{1}^{\dagger}-2{\;\;\!\!\!\rm cos}(\theta+\varphi)\hat{a}_{1}\hat{\rho}(t)\hat{a}_{3}^{\dagger}\Big\}.
		\label{A12}
	\end{align}
\end{widetext}
After neglecting the effects of terms $\hat{a}_{j}\hat{\rho}(t)\hat{a}_{k}^{\dagger}$ representing quantum jumps, where $j$ and $k=1,2,3$, we find that the evolution of the system can be described by the effective non-Hermitian Hamiltonian $\hat{H}_{\rm NH}$. The specific evolution equation takes the following form
\begin{equation}
	i\dfrac{d\hat{\rho}(t)}{dt}=\Big[\hat{H}_{\rm NH}\hat{\rho}(t)-\hat{\rho}(t)\hat{H}_{\rm NH}^{\dagger}\Big]\;\!\!,
\end{equation}
in which
\begin{widetext}
	\begin{align}
\hat{H}_{\rm NH}&=\hat{H}_{\rm OM}-i\sum_{j=1}^{3}2\pi J_{j}^{2}\;\;\!\!\!\omega_{0}\hat{a}_{j}^{\dag}\hat{a}_{j}-i2\pi J_{1}J_{2}\;\;\!\!\!\omega_{0}(\hat{a}_{2}^{\dagger}\hat{a}_{1}e^{i\theta}+\hat{a}_{1}^{\dagger}\hat{a}_{2}e^{i\theta}) \nonumber \\
&-i2\pi J_{2}J_{3}\;\;\!\!\!\omega_{0}(\hat{a}_{3}^{\dagger}\hat{a}_{2}e^{i\varphi}+\hat{a}_{2}^{\dagger}\hat{a}_{3}e^{i\varphi})-i2\pi J_{1}J_{3}\;\;\!\!\!\omega_{0}\Big[\hat{a}_{3}^{\dagger}\hat{a}_{1}e^{i(\theta+\varphi)}+\hat{a}_{1}^{\dagger}\hat{a}_{3}e^{i(\theta+\varphi)}\Big]\;\!\!.
\label{A14}
	\end{align}
\end{widetext}
It is obvious that the common environment shared by the subsystems will cause additional dissipation to the entire system, that is
\begin{equation}
\hat{H}_{\rm D}=-\:i\sum_{j=1}^{3}2\pi J_{j}^{2}\;\;\!\!\!\omega_{0}\hat{a}_{j}^{\dag}\hat{a}_{j},
\end{equation}
which is similar to the independent environment. However, the difference is that the common environment will induce interactions between subsystems (i.e., microwave optomechanical resonators). The effective non-Hermitian interaction Hamiltonian between resonators mediated by the common environment is
\begin{widetext}
	\begin{align}
		\hat{H}_{\rm NI}=-\:i2\pi J_{1}J_{2}\;\;\!\!\!\omega_{0}(\hat{a}_{2}^{\dagger}\hat{a}_{1}e^{i\theta}+\hat{a}_{1}^{\dagger}\hat{a}_{2}e^{i\theta})-i2\pi J_{2}J_{3}\;\;\!\!\!\omega_{0}(\hat{a}_{3}^{\dagger}\hat{a}_{2}e^{i\varphi}+\hat{a}_{2}^{\dagger}\hat{a}_{3}e^{i\varphi})-i2\pi J_{1}J_{3}\;\;\!\!\!\omega_{0}\Big[\hat{a}_{3}^{\dagger}\hat{a}_{1}e^{i(\theta+\varphi)}+\hat{a}_{1}^{\dagger}\hat{a}_{3}e^{i(\theta+\varphi)}\Big]\;\!\!.
	\end{align}
\end{widetext}
Therefore, the influence of the common environment on microwave optomechanical circuits can be equivalently described by the non-Hermitian Hamiltonian $\hat{H}_{\rm N}=\hat{H}_{\rm D}+\hat{H}_{\rm NI}$, i.e., Eq.~(\ref{Eq4}) in the main text.

\section{equations of motion for the microwave optomechanical resonators \label{Appendix2}}

In this appendix, we provide the detailed procedure for deriving the equations of motion Eqs.~(\ref{Eq6})--(\ref{Eq11}). We note that for the microwave optomechanical resonators shown in Eqs.~(\ref{Eq1})--(\ref{Eq3}), the common environment will induce an additional dissipation to each resonator and induce non-Hermitian interactions between the microwave modes of any two resonators, as demonstrated in Eq.~(\ref{A14}). If we apply an ac current with frequency $\omega_{0}$ and amplitude $\varepsilon$ to each microwave mode, then the Heisenberg-Langevin equations of motion for the microwave and mechanical modes can be written as
\begin{align}
	\dfrac{d\hat{a}_{j}}{dt}&=\left[-\:i\omega_{j}-\gamma_{j}+iG(\hat{b}_{j}^{\dag}+\hat{b}_{j})\right]\!\hat{a}_{j}+\varepsilon e^{-i\omega_{0}t} \nonumber \\&-i\!\int_{0}^{\infty}\!d\omega J_{j}
	\sqrt{\omega}\left[\hat{c}_{\rm L}(\omega)e^{-i\omega x_{j}/v}+\hat{c}_{\rm R}(\omega)e^{i\omega x_{j}/v}\right]\!\;\!\!, \nonumber\\	\dfrac{d\hat{b}_{j}}{dt}&=(-\:i\Omega_{j}-\Gamma_{j})\:\!\hat{b}_{j}+iG\hat{a}_{j}^{\dag}\hat{a}_{j},
	\label{B1}
\end{align}	
in which all the parameters and operators are the same as those defined in Appendix~\ref{Appendix1}. Furthermore, the common environmental modes $\hat{c}_{\rm L}(\omega)$ and $\hat{c}_{\rm R}(\omega)$ should satisfy the following equations
\begin{equation}
	\begin{split}
		\dfrac{d\hat{c}_{\rm L}(\omega)}{dt}=&-i\omega\hat{c}_{\rm L}(\omega)-i\sum_{j=1}^{3} J_{j}\sqrt{\omega}\;\;\!\!\!\hat{a}_{j}e^{i\omega x_{j}/v},\\
		\dfrac{d\hat{c}_{\rm R}(\omega)}{dt}=&-i\omega\hat{c}_{\rm R}(\omega)-i\sum_{j=1}^{3} J_{j}\sqrt{\omega}\;\;\!\!\!\hat{a}_{j}e^{-i\omega x_{j}/v}.
		\label{B2}
	\end{split}
\end{equation}
By utilizing the result in Eq.~(\ref{A11}), we can eliminate the common environmental modes and obtain
\begin{align}
	\dfrac{d\hat{a}_{1}}{dt}=&\left[-\:i\omega_{1}-\gamma_{1}-2\pi J_{1}^2\omega_{0}+iG(\hat{b}_{1}+\hat{b}_{1}^\dagger)\right]\!\hat{a}_{1}+\varepsilon e^{-i\omega_{0}t} \nonumber\\
	-&\:2\pi J_{1}J_{2}\;\;\!\!\!\omega_{0}e^{i\theta}\hat{a}_{2}-2\pi J_{1}J_{3}\;\;\!\!\!\omega_{0}e^{i(\theta+\varphi)}\hat{a}_{3}, \label{B3}\\
	\dfrac{d\hat{a}_{2}}{dt}=&\left[-\:i\omega_{2}-\gamma_{2}-2\pi J_{2}^2\omega_{0}+iG(\hat{b}_{2}+\hat{b}_{2}^\dagger)\right]\!\hat{a}_{2}+\varepsilon e^{-i\omega_{0}t} \nonumber\\
	-&\:2\pi J_{1}J_{2}\;\;\!\!\!\omega_{0}e^{i\theta}\hat{a}_{1}-2\pi J_{2}J_{3}\;\;\!\!\!\omega_{0}e^{i\varphi}\hat{a}_{3},\\
	\dfrac{d\hat{a}_{3}}{dt}=&\left[-\:i\omega_{3}-\gamma_{3}-2\pi J_{3}^2\omega_{0}+iG(\hat{b}_{3}+\hat{b}_{3}^\dagger)\right]\!\hat{a}_{3}+\varepsilon e^{-i\omega_{0}t} \nonumber\\-&\:2\pi J_{1}J_{3}\;\;\!\!\!\omega_{0}e^{i(\theta+\varphi)}\hat{a}_{1}-2\pi J_{2}J_{3}\;\;\!\!\!\omega_{0}e^{i\varphi}\hat{a}_{2},\\
	\dfrac{d\hat{b}_{j}}{dt}=&\:(-\:i\Omega_{j}-\Gamma_{j})\:\!\hat{b}_{j}+iG\hat{a}_{j}^{\dag}\hat{a}_{j}, \quad j=1,2,3,
	\label{B6}
\end{align}
in which $\gamma_{j}$ and $\Gamma_{j}$ represent the intrinsic damping rates (the independent environmental effect) of the microwave and mechanical modes of the $j$-th resonator, respectively. Here, the fluctuation operators have been neglected since they have no effect on the expectation value equations of the system operators. Moreover, we define
\begin{equation}
a_{j}e^{-i\omega_{0}t}=\langle\hat{a}_{j}\rangle,\ a_{j}^{*}e^{i\omega_{0}t}=\langle\hat{a}_{j}^{\dagger}\rangle,\
b_{j}=\langle\hat{b}_{j}\rangle,\
b_{j}^{*}=\langle\hat{b}_{j}^{\dagger}\rangle,
\end{equation}
where $\langle\hat{o}\rangle$ represents the expectation value of any system operator $\hat{o}$. Moreover, we define the parameters as $\omega_{1}=\omega_{0}-\delta$, $\omega_{2}=\omega_{0}$, and $\omega_{3}=\omega_{0}+\delta$. Then Eqs.~(\ref{B3})--(\ref{B6}) can be rewritten as
\begin{align}
	\dfrac{da_{1}}{dt}=&\left[i\delta-\gamma_{1}-2\pi J_{1}^2\omega_{0}+iG(b_{1}+b_{1}^{*})\right]\!a_{1}+\varepsilon \nonumber\\
	-&\:2\pi J_{1}J_{2}\;\;\!\!\!\omega_{0}e^{i\theta}a_{2}-2\pi J_{1}J_{3}\;\;\!\!\!\omega_{0}e^{i(\theta+\varphi)}a_{3}, \\
	\dfrac{da_{2}}{dt}=&\left[-\:\gamma_{2}-2\pi J_{2}^2\omega_{0}+iG(b_{2}+b_{2}^{*})\right]\!a_{2}+\varepsilon \nonumber\\
	-&\:2\pi J_{1}J_{2}\;\;\!\!\!\omega_{0}e^{i\theta}a_{1}-2\pi J_{2}J_{3}\;\;\!\!\!\omega_{0}e^{i\varphi}a_{3}, \\
	\dfrac{da_{3}}{dt}=&\left[-\:i\delta-\gamma_{3}-2\pi J_{3}^2\omega_{0}+iG(b_{3}+b_{3}^{*})\right]\!a_{3}+\varepsilon \nonumber\\-&\:2\pi J_{1}J_{3}\;\;\!\!\!\omega_{0}e^{i(\theta+\varphi)}a_{1}-2\pi J_{2}J_{3}\;\;\!\!\!\omega_{0}e^{i\varphi}a_{2}, \\
    \dfrac{db_{j}}{dt}=&\:(-\:i\Omega_{j}-\Gamma_{j})\:\!b_{j}+iG|a_{j}|^2, \quad j=1,2,3.
\end{align}
We have applied the mean-field approximation~\cite{PRL111073603}, and neglected the second- and higher-order correlations between the microwave-microwave modes and the microwave-mechanical modes~\cite{PRL103213603}.

\end{document}